\documentclass[nofootinbib, tightenlines]{revtex4}

\usepackage{axodraw}
\usepackage{graphicx}

\def\la{\langle}
\def\ra{\rangle}

\def\beq{\begin{equation}}
\def\eeq{\end{equation}}
\def\bea{\begin{eqnarray}}
\def\eea{\end{eqnarray}}
\def\barr{\begin{array}}
\def\earr{\end{array}}

\def\mev{\mbox{ MeV}}
\def\gev{\mbox{ GeV}}

\def\op{{\mathcal{O}}}
\def\ampl{{\mathcal{M}}}

\def\zsum{{\mathcal{H}}}
\def\etaprime{\eta^{\prime}}
\def\exponential{{\mathrm{e}}}
\def\calC{{\mathcal{C}}}

\begin{document}

\begin{titlepage}
\begin{flushright}
INT-PUB 04-05\\
NT@UW-04-03\\
UW/PT 04-02\\
\end{flushright}
\vskip 0.5cm
\begin{center}
{\Large \bf Heavy meson chiral perturbation theory in finite volume\\} 
\vskip1cm {\large\bf
Daniel~Arndt$^{a}$ and C.-J. David~Lin$^{a,b}$}\\ \vspace{.5cm}
{\normalsize {\sl $^a$ Department of Physics,  
University of Washington, Seattle, WA 98195-1560, USA.\\
$^{b}$ Institute for Nuclear Theory, 
University of Washington, Seattle, WA 98195-1550, USA.}}

\vskip1.0cm {\large\bf Abstract:\\[10pt]} \parbox[t]{\textwidth}{{
We study finite volume effects in heavy quark systems 
in the framework of heavy meson chiral perturbation theory for
full, quenched, and partially quenched QCD.
A novel feature of this investigation is the role played by the 
scales $\Delta_{\ast}$ and ${\delta}_{s}$, where  
$\Delta_{\ast}$ is the mass difference
between the heavy-light vector and pseudoscalar mesons of the
same quark content, and ${\delta}_{s}$ is the mass difference 
due to light flavour $SU(3)$ breaking.  The 
primary conclusion of this work is that finite
volume effects arising from the propagation of Goldstone particles in 
the effective theory
can be altered by the presence of these scales.
Since $\Delta_{\ast}$ varies significantly 
with the heavy quark mass, these volume effects can be amplified
in both heavy and light quark mass extrapolations (interpolations).
As an explicit example, we present 
results for $B$ parameters of neutral $B$ meson mixing matrix elements
and heavy-light decay constants to
one-loop order in finite volume heavy meson chiral perturbation theory for
full, quenched, and $N_{f}=2+1$ partially quenched QCD.
Our calculation shows that for 
high-precision
determinations of the phenomenologically interesting $SU(3)$ breaking 
ratios, finite volume effects are significant in 
quenched and not negligible in partially quenched QCD, 
although they are generally small in full QCD.}}
\end{center}
\end{titlepage}

\section{Introduction}
\label{sec:intro}
Numerical calculations of hadronic properties using lattice QCD have provided
significant inputs to particle physics phenomenology.  In particular, the
joint effort between experiment and theory to investigate the unitarity
triangle in the Cabibbo-Kobayashi-Maskawa (CKM) matrix from $B$ meson
decays and mixing has made
impressive progress~\cite{Battaglia:2003in}, in which lattice QCD has played 
an important role.  Nevertheless, current lattice calculations are 
still subject to various systematic errors.  In this paper, we address 
finite volume effects which arise 
in lattice calculations for heavy-light meson systems
from the light degrees of freedom.  Our framework is
heavy meson chiral perturbation theory (HM$\chi$PT) with first 
order $1/M_{P}$ and chiral corrections.
We assume the mass hierarchy
\beq
\label{eq:mass_hierarchy}
 M_{\mathrm{GP}} \ll \Lambda_{\chi} \ll M_{P} ,
\eeq
where $M_{\mathrm{GP}}$ is the mass of any Goldstone particle, 
$M_{P}$ is the mass of the heavy-light meson, and 
$\Lambda_{\chi}$ is the chiral symmetry breaking scale.  Under this
assumption, we discard corrections of the size
\beq
\label{eq:mgp_over_mp}
 \frac{M_{\mathrm{GP}}}{M_{P}} .
\eeq
Concerning the finite volume, we work with the
condition that
\beq\label{eq:MpiL_big}
 M_{\mathrm{GP}} L \gg 1,
\eeq
where $L$ is the spatial extent of the cubic box.  Therefore,
given that $f_{\pi} L/\sqrt{2}$ ($f_{\pi} \approx 132 \mev$) will be close to 
one in lattice simulations in the near future, 
one can still neglect the chiral symmetry restoration effects resulting from
the Goldstone zero momentum modes~\cite{Gasser:1987ah,Leutwyler:1992yt} 
when Eq.~(\ref{eq:MpiL_big}) is satisfied.

%
%

The main task of this work is to study the volume effects due to the presence
of the scales 
\beq
\label{eq:Delta_def}
 \Delta_{\ast} = M_{P^{\ast}} - M_{P} ,
\eeq
and
\beq
\label{eq:delta_s_def}
 \delta_{s} = M_{P_{s}} - M_{P} ,
\eeq
where $P^{\ast}$ and $P$ are the heavy-light vector and pseudoscalar mesons 
containing a $u$ or $d$ anti-quark\footnote{We work in the isospin limit in
this paper.}, and $P_{s}$ is the 
heavy-light pseudoscalar meson with an $s$ anti-quark. The
scale $\Delta_{\ast}$ appears due to the breaking of heavy quark spin symmetry
that is of $\op(1/M_{P})$ and $\delta_{s}$ comes from light flavour $SU(3)$ 
breaking in the heavy-light meson masses.  Under the assumption
of Eq.~(\ref{eq:mass_hierarchy}), $\Delta_{\ast}$ is independent of the
light quark mass, and $\delta_{s}$ does not contain any $1/M_{P}$
corrections, at the order
we are working.

%
%

In the real world, both $\Delta_{\ast}$ and $\delta_{s}$ are not very 
different from the pion mass.  In fact~\cite{Hagiwara:2002fs},
\beq
 M_{B_{s}} - M_{B} =  91 \mev,
\eeq
\beq
 M_{D_{s}} - M_{D} =  104 \mev,
\eeq
\beq
 M_{B^{\ast}} - M_{B} = 46 \mev,
\eeq
and
\beq
 M_{D^{\ast}} - M_{D} = 142 \mev .
\eeq
In current lattice simulations, these mass splittings 
vary between 0 and $\sim 150$
MeV.  Therefore it is important to include them
in the investigation of finite volume effects.
Equation~(\ref{eq:MpiL_big}) implies that the Compton
wavelength of the Goldstone particle is small compared
to the size of the box.  Therefore
finite volume effects mainly result from the propagation of 
the Goldstone particles
to the boundary.
However, as shown in Section \ref{sec:FV}, 
$\Delta_{\ast}$ and $\delta_{s}$ can, 
in a non-trivial way, alter these 
effects.  In particular, since $\Delta_{\ast}$ varies 
with the heavy quark mass,
finite volume effects can be significantly amplified 
in heavy quark mass extrapolations.

%
%

This paper is organised as follows.  In Section \ref{sec:HMChPT} we summarise
the ingredients of HM$\chi$PT relevant to this work.  Section
\ref{sec:FV} is devoted to the discussion of HM$\chi$PT in finite volume,
emphasising the role played by $\delta_{s}$ and
$\Delta_{\ast}$.  
We then present an explicit
calculation of neutral $B$ meson mixing and heavy-light decay constants in 
Section \ref{sec:BBbar_mixing} and discuss the phenomenological impact that
finite volume effects can have.  We conclude in 
Section \ref{sec:conclusion}.   
Some mathematical formulae and results are summarised in the appendices.  

%
%

As this work progressed, we were informed that similar ideas and techniques
were also being applied in heavy baryon chiral perturbation theory 
\cite{beane:to_appear}\footnote{We thank Silas Beane for drawing our attention 
to his work.}.  
Although the
underlying physics is somewhat different, many technical 
aspects are quite
similar to those presented here.

\section{Heavy meson chiral perturbation theory}
\label{sec:HMChPT}
The chiral Lagrangian for the Goldstone particles is
\bea
 {\mathcal{L}}_{\mathrm{GP}} &=&
 \frac{f^{2}}{8} {\mathrm{(s)tr}} \left [ 
 \big (\partial_{\mu} \Sigma^{\dagger}\big )\big (\partial^{\mu} \Sigma \big)
 + \Sigma^{\dagger} \chi + \chi^{\dagger} \Sigma\right ]
     \nonumber \\
 &&+ A_{\eta^{\prime}} \left [
\alpha (\partial_{\mu}\Phi_{0})(\partial^{\mu} \Phi_{0})
  - M_{0}^{2} \Phi^{2}_{0} 
        \right ] ,
\eea
where $A_{\eta^{\prime}}=1$ for quenched QCD (QQCD) and 
partially quenched QCD (PQQCD), and $A_{\eta^{\prime}}=0$ for full QCD.
$\Sigma= {\mathrm{exp}}(2 i \Phi/f)$ is the non-linear Goldstone particle
field, with
$\Phi$ being the matrix containing the standard Goldstone 
fields.%
\footnote{In this paper, we only address situations where there are 
no multi-particle thresholds involved in loops.
This is the case for the explicit calculation presented in
Section~\ref{sec:BBbar_mixing}.  
Therefore, in spite of the sickness pointed out in 
Ref.~\cite{Bernard:1996ez}, we can still
use the Minkowski formalism even for the case of (P)QQCD. 
This makes the physics discussion in Section~\ref{sec:FV}
simpler compared to the Euclidean formalism.
The effects from multi-particle thresholds in finite volume
HM$\chi$PT will be discussed in a future publication.} 
We use $f=132$~MeV. 
In this work, we
follow the supersymmetric formulation of (partially) quenched chiral
perturbation theory
[(P)Q$\chi$PT] \cite{Bernard:1992mk,Bernard:1994sv}.
Therefore $\Sigma$ transforms linearly under 
$SU(3)_{\mathrm{L}}\otimes SU(3)_{\mathrm{R}}$,
$SU(3|3)_{\mathrm{L}}\otimes SU(3|3)_{\mathrm{R}}$ and 
$SU(6|3)_{\mathrm{L}}\otimes SU(6|3)_{\mathrm{R}}$ in full QCD, QQCD and 
PQQCD respectively.  The symbol ``(s)tr'' in the above equation
means ``trace'' in full QCD and ``supertrace'' in (P)QQCD.
The variable $\chi$ is defined as
\beq
 \chi \equiv 2 B_{0} {\mathcal{M}}_{q} = 
\frac{-2 \la 0 |\bar{u}u+\bar{d}d|0\ra}{f^{2}}
 {\mathcal{M}}_{q},
\eeq
where the quark mass matrix $\ampl_{q}$ is
\beq
\label{eq:full_mass_matrix}
 \ampl^{\mathrm{(QCD)}}_{q} = {\mathrm{diag}} 
 (m,m,m_{s}) ,
\eeq
in full QCD,
\beq
\label{eq:quenched_mass_matrix}
 \ampl^{({\mathrm{QQCD}})}_{q} = {\mathrm{diag}} 
 (\underbrace{m,m,m_{s}}_{{\mathrm{valence}}},
 \underbrace{m,m,m_{s}}_{{\mathrm{ghost}}}) ,
\eeq
in QQCD, and 
\beq
\label{eq:PQ_mass_matrix}
 \ampl^{({\mathrm{PQQCD}})}_{q} = {\mathrm{diag}} 
 (\underbrace{m,m,m_{s}}_{{\mathrm{valence}}},
  \underbrace{\tilde{m},\tilde{m},\tilde{m}_{s}}_{{\mathrm{sea}}},
  \underbrace{m,m,m_{s}}_{{\mathrm{ghost}}}) ,
\eeq
in PQQCD.
We keep the strange quark mass different from that 
of the up and down quarks
in the valence, sea and ghost sectors.  
Notice that the flavour singlet state $\Phi_0={\mathrm{str}}(\Phi)/\sqrt{6}$ 
is rendered heavy by the
$U(1)_A$ anomaly in PQQCD \cite{Sharpe:2001fh,Sharpe:2000bc}
and can be integrated out;
it has to be kept as a dynamical degree
of freedom in QQCD.

%
%

The inclusion of the heavy-light mesons in 
chiral perturbation theory (HM$\chi$PT) was first proposed in
Refs.~\cite{Burdman:1992gh, Wise:1992hn,Yan:1992gz}, with the
generalisation to 
quenched and partially quenched theories given in 
Refs.~\cite{Booth:1995hx, Sharpe:1996qp}.  The $1/M_{P}$ and
chiral corrections were studied by Boyd and Grinstein 
\cite{Boyd:1995pa} in full QCD and by Booth \cite{Booth:1994rr}
in QQCD.  The spinor field appearing in this effective theory is 
\beq
\label{eq:H_field}
 H^{(Q)}_{a} = \frac{1 + \slash\!\!\! v}{2} \left ( 
 P^{\ast (Q)}_{a,\mu} \gamma^{\mu} - P^{(Q)}_{a}\gamma_{5}\right ) ,
\eeq
where $P^{(Q)}_{a}$ and $P^{\ast (Q)}_{a,\mu}$ annihilate 
pseudoscalar and vector
mesons containing a heavy quark $Q$ and a light anti-quark of flavour $a$.  
Under a heavy quark spin $SU(2)$ transformation $S$, 
\beq
 H^{(Q)}_{a} \longrightarrow S H^{(Q)}_{a} .
\eeq
Under the vector light-flavour transformation $U$ [{\it i.e.}, $U\in SU(3)$ 
for full QCD,
$U\in SU(3|3)$ for QQCD and $U\in SU(6|3)$ for PQQCD], 
\beq
 H^{(Q)}_{a} \longrightarrow H^{(Q)}_{b}
  U^{\dagger}_{ba}.
\eeq
Also, the conjugate field, which creates heavy-light mesons containing a heavy
quark $Q$ and a light anti-quark of flavour $a$, is defined as
\beq
 \bar{H}^{(Q)}_{a} = \gamma^{0} H^{(Q)} \gamma_{0}.
\eeq
Furthermore, the Goldstone particles appear in the HM$\chi$PT Lagrangian 
via the field
\beq
 \xi \equiv \exponential^{i\Phi/f} ,
\eeq
which transforms as
\beq
 \xi \longrightarrow U_{\mathrm{L}} 
  \xi U^{\dagger} = U \xi 
   U^{\dagger}_{\mathrm{R}},
\eeq
where $U_{\mathrm{L(R)}}$ is an element of the left-handed (right-handed) 
$SU(3)$, $SU(3|3)$ and $SU(6|3)$ groups for QCD, QQCD, and PQQCD respectively.
The HM$\chi$PT Lagrangian, to lowest order in the chiral and
$1/M_{Q}$ expansion, for mesons containing a heavy quark $Q$ and a
light anti-quark of flavour $a$ is then
%
\bea
\label{eq:HMChPT}
 {\mathcal{L}}_{\mathrm{HM\chi PT}} 
 &=& -i \, {\mathrm{tr_{D}}}
  \left (
   \bar{H}^{(Q)}_{a} v_{\mu} \partial^{\mu} H^{(Q)}_{a}
  \right )
+ \frac{i}{2}  {\mathrm{tr_{D}}}
  \left (
   \bar{H}^{(Q)}_{a} v_{\mu} \left [ \xi^{\dagger}\partial^{\mu}
   \xi + \xi
  \partial^{\mu}\xi^{\dagger}\right ]_{ab} H^{(Q)}_{b}
  \right )
 \nonumber\\
& & + \frac{i}{2} g\,
   {\mathrm{tr_{D}}}
  \left (
  \bar{H}^{(Q)}_{a} \gamma_{\mu}\gamma_{5}
   \left [ \xi^{\dagger}\partial^{\mu}\xi 
        - \xi\partial^{\mu}
  \xi^{\dagger}\right ]_{ab}   
  H^{(Q)}_{b}
  \right )
+ B_{\etaprime}\frac{i}{2}\gamma\,
 {\mathrm{tr_{D}}}
  \left (
  \bar{H}^{(Q)}_{a}  H^{(Q)}_{a}\gamma_{\mu}\gamma_{5}
  \right )
  {\mathrm{str}} \left [ 
  \xi^{\dagger}\partial^{\mu}\xi 
   - \xi\partial^{\mu}\xi^{\dagger}\right ] ,
    \nonumber \\
\eea
%
where $B_{\etaprime}=0$ for full QCD, and $B_{\etaprime}=1$
for (P)QQCD\footnote{However, since we integrate out the $\eta^{\prime}$
in PQQCD, the coupling $\gamma$ does not appear in the results presented
in this paper.}.    
We do not distinguish the coupling $g$ in these theories.
It is implicitly assumed that 
``${\mathrm{(s)tr}}_{a}$'' is taken appropriately in flavour space. 
${\mathrm{tr_{D}}}$ means taking the trace in
Dirac space.  
The HM$\chi$PT Lagrangian for mesons containing a 
heavy anti-quark $\bar{Q}$ and a light quark of flavour $a$ is obtained
by applying the charge conjugation operation to the above Lagrangian
\cite{Grinstein:1992qt}.
At this order, 
the propagators for $P^{(Q)}_{a}$ and $P^{\ast (Q)}_{a}$ mesons are
\beq
 \frac{i}{2 (v\cdot k + i\epsilon)} \mbox{ },\mbox{ }\mbox{ }
 \frac{-i (g_{\mu\nu} - v_{\mu}v_{\nu})}{2 (v\cdot k + i\epsilon)} ,
\eeq
respectively.

%

The effects of chiral and heavy quark symmetry breaking have been 
systematically studied in full~\cite{Boyd:1995pa} and
quenched~\cite{Booth:1994rr} HM$\chi$PT.  Amongst them, the only 
relevant feature necessary for the purpose of this work, {\it i.e.},
the investigation of finite volume effects, 
are the shifts to the masses of the heavy-light mesons.
These shifts are from the heavy quark spin breaking term
\beq
\label{eq:HQ_spin_breaking_term}
\frac{\lambda_{2}}{M_{P}}
{\mathrm{tr_{D}}} \left (
 \bar{H}^{(Q)}_{a}\sigma_{\mu\nu} H^{(Q)}_{a} \sigma^{\mu\nu}
\right ) ,
\eeq
and the chiral symmetry breaking terms
\bea
\label{eq:su3_violating_terms}
 &&\lambda_{1}B_{0}\mbox{ }{\mathrm{tr_{D}}} \left ( 
  \bar{H}^{(Q)}_{a} 
  \left [
   \xi {\mathcal{M}}_{q} \xi + 
   \xi^{\dagger} {\mathcal{M}}_{q} \xi^{\dagger}
  \right ]_{ab}
  H^{(Q)}_{b}
 \right )
\nonumber \\
&&+ \lambda_{1}^{\prime}B_{0}\mbox{ }{\mathrm{tr_{D}}} \left ( 
  \bar{H}^{(Q)}_{a} 
  H^{(Q)}_{a}
 \right )
 \left [
   \xi {\mathcal{M}}_{q} \xi + 
   \xi^{\dagger} {\mathcal{M}}_{q} \xi^{\dagger}
  \right ]_{bb} .
\nonumber\\
\eea
We choose to work with the effective theory in which the 
heavy-light pseudoscalar
mesons that contain a heavy quark and a $u$ or $d$ valence 
anti-quark are massless.
Notice that the term proportional to $\lambda^{\prime}_{1}$ in 
Eq.~(\ref{eq:su3_violating_terms}) causes a universal shift to
all the heavy-light meson masses.
This means that the
masses appearing in the propagators of heavy vector mesons and 
any meson containing an $s$ anti-quark (valence or ghost) are shifted 
as follows:
\beq
\label{eq:shift_start}
  \frac{-i (g_{\mu\nu} - v_{\mu} v_{\nu})}
   {2 (v\cdot k - \Delta_{\ast} + i\epsilon)},\mbox{ }\mbox{ }
  \frac{i}{2 (v\cdot k - \delta_{s} + i\epsilon)},
\eeq
and
\beq
  \frac{-i (g_{\mu\nu} - v_{\mu} v_{\nu})}
   {2 (v\cdot k - \Delta_{\ast} - \delta_{s} + i\epsilon)} ,
\eeq
for $P^{\ast}$, $P_{s}$, and $P^{\ast}_{s}$ (heavy vector meson containing an
$s$ valence or ghost anti-quark), respectively.  The mass
shifts can be written in terms of the couplings in 
Eqs.~(\ref{eq:HQ_spin_breaking_term}) and  
(\ref{eq:su3_violating_terms}):
\beq
 \Delta_{\ast} = -8 \frac{\lambda_{2}}{M_{P}} ,
\eeq
and
\beq
\label{eq:delta_s_lambda}
 \delta_{s} = 2\lambda_{1} B_{0} (m_{s}-m) .
\eeq

In PQQCD, there are two additional mass shifts
because the
sea quarks have different masses from the valence and ghost quarks:
\beq
\label{eq:tilde_delta_s_def}
 \tilde{\delta}_{s} = M_{\tilde{P}_{s}} - M_{\tilde{P}}
      = 2\lambda_{1} B_{0} (\tilde{m}_{s}-\tilde{m}) ,
\eeq
and
\beq
\label{eq:delta_sea_def}
 \delta_{\mathrm{sea}} = M_{\tilde{P}} - M_{P}
 = 2\lambda_{1} B_{0} (\tilde{m}-m) .
\eeq
where $\tilde{P}$ ($\tilde{P}_{s}$) is
the heavy-light pseudoscalar meson with a $d$ ($s$) sea 
anti-quark.  The propagators of the heavy mesons containing 
sea anti-quarks are:
\beq
  \frac{i}
   {2 (v\cdot k - \delta_{\mathrm{sea}} + i\epsilon)},
\eeq
\beq
  \frac{-i (g_{\mu\nu} - v_{\mu} v_{\nu})}
   {2 (v\cdot k - \Delta_{\ast} - \delta_{\mathrm{sea}} 
  + i\epsilon)}, 
\eeq
\beq
  \frac{i}{2 (v\cdot k - \delta_{\mathrm{sea}} - \tilde{\delta}_{s} 
   + i\epsilon)},
\eeq
and
\beq
\label{eq:shift_end}
  \frac{-i (g_{\mu\nu} - v_{\mu} v_{\nu})}
   {2 (v\cdot k  - \Delta_{\ast} 
    - \delta_{\mathrm{sea}} - \tilde{\delta}_{s} + i\epsilon)}
\eeq
for $\tilde{P}$, $\tilde{P}^*$ (vector meson with a $d$ sea anti-quark), 
$\tilde{P}_s$, 
and $\tilde{P}^*_s$ (vector meson with an $s$ sea anti-quark), respectively.

\section{Finite volume effects}
\label{sec:FV}
In this section, we discuss generic features of finite volume effects
in HM$\chi$PT.  For clarity, we use the symbol $\Delta$ for one of 
($\Delta_{\ast}$, $\delta_{s}$, $\tilde{\delta}_{s}$, 
$\delta_{\mathrm{sea}}$) or any
sum amongst them.

%
%


In the limit where the heavy quark mass goes to infinity and the light quark
masses are equal, all the heavy mesons in HM$\chi$PT become on-shell 
static sources, and there is a velocity superselection rule when the
momentum transfer involved in the scattering of the heavy meson system
is fixed~\cite{Georgi:1990um}.  For illustration, consider the 
vertex with coupling $g$ in 
${\mathcal{L}}_{\mathrm{HM\chi PT}}$ introduced in 
Eq.~(\ref{eq:HMChPT}).  The heavy-light meson
$P$ can scatter into $P^{\ast}_{(s)}$ by emitting a Goldstone particle with
mass $M_{\mathrm{GP}}$ through this vertex.  
The momenta of the mesons $P$ and $P^{\ast}_{(s)}$
are
\beq
 M_{P}v_{\mu} ,
\eeq
and
\beq
 M_{P^{\ast}_{(s)}}v_{\mu} + k_{\mu} = M_{P}v_{\mu} + k_{\mu} ,
\eeq
where the velocity 
$v_{\mu}=(1,0,0,0)$ in the rest frame of the heavy mesons, and 
$k_{\mu}$ is the soft momentum carried by the Goldstone particle.
The infinitely heavy $P$ and $P^{\ast}_{(s)}$ mesons
do not propagate in space.
Therefore, when such a system is in
a cubic spatial box, 
finite volume effects result entirely from the
propagation of the Goldstone particle to the boundary with momentum
$k\sim M_{\mathrm{GP}}$. In this case, the volume effects behave like
${\mathrm{exp}}(-M_{\mathrm{GP}}L)$ multiplied by a polynomial in
$1/L$.

%
%

The breaking of heavy quark spin and $SU(3)$ light flavour symmetries 
in HM$\chi$PT can induce a mass difference
\beq
 M_{P^{\ast}_{(s)}} = M_{P} + \Delta ,
\eeq
which complicates the above picture.  
In this scenario,
the $P^{\ast}_{(s)}$ is still regarded 
as a static source, but it is off-shell with the virtuality $\Delta$.
The period during which the Goldstone particle can propagate to the
boundary is limited by the time uncertainty conjugate to
this virtuality, {\it i.e.},
\beq
\label{eq:virtuality}
\delta t \sim \frac{1}{\Delta} .
\eeq
This means that finite volume effects, which arise from the propagation of the 
Goldstone particles in such a system, will decrease as $\Delta$ increases. 
Equation (\ref{eq:virtuality}) also indicates that the suppression of the 
volume effects by a non-zero $\Delta$ is controlled by the parameter
\beq
\label{eq:mass_ratio}
 \frac{M_{\mathrm{GP}}}{\Delta} .
\eeq
%

%
%

To see explicitly how this phenomenon appears in a calculation,
we consider a typical sum in one-loop HM$\chi$PT, with a 
Goldstone propagator and a heavy-light vector meson propagator in 
the loop, in a cubic box with periodic boundary condition:
%
\beq
\label{eq:calJ}
 {\mathcal{J}}(M_{\mathrm{GP}},\Delta) 
 = -i  \frac{1}{L^{3}}
 \sum_{\vec{k}} \int \frac{d k_{0}}{2\pi}
 \frac{1}{(k^{2}-M^{2}_{\mathrm{GP}} + i\epsilon )
 (v\cdot k - \Delta + i\epsilon )}  ,
\eeq
where the spatial momentum $\vec{k}$ is quantised in finite 
volume as
\beq
\label{eq:k_quantum}
 \vec{k} = \left (\frac{2\pi}{L}\right ) \vec{i} ,
\eeq
with $\vec{i}$ being a three dimensional integer vector.
Using the Poisson summation formula, it is straightforward to show that
\beq
  {\mathcal{J}}(M_{\mathrm{GP}},\Delta) 
 = J(M_{\mathrm{GP}},\Delta)
 + J_{\mathrm{FV}}(M_{\mathrm{GP}},\Delta),
\eeq
where 
\beq
\label{eq:J}
 J(M_{\mathrm{GP}},\Delta) = -i \int \frac{d^{4}k}{(2\pi)^{4}}
 \frac{1}{(k^{2}-M^{2}_{\mathrm{GP}} +i\epsilon )
 (v\cdot k - \Delta + i\epsilon )}  ,
\eeq
is the infinite volume limit of 
${\mathcal{J}}(M_{\mathrm{GP}},\Delta)$, and  
\beq
\label{eq:JFV}
 J_{\mathrm{FV}}(M_{\mathrm{GP}},\Delta) =
 \left (\frac{1}{2\pi} \right )^{2} 
 \sum_{\vec{n}\not = \vec{0}} \int_{0}^{\infty}
 d |\vec{k}| \left ( \frac{|\vec{k}|}{w_{\vec{k}}\mbox{ }
 (w_{\vec{k}} + \Delta)} \right )
 \left ( \frac{{\mathrm{sin}}(|\vec{k}| |\vec{n}| L)}{|\vec{n}| L}\right ) ,
\eeq
with
\beq
 w_{\vec{k}} = \sqrt{|\vec{k}|^{2}+M^{2}_{\mathrm{GP}}} ,
\eeq
is the finite volume correction to $J(M_{\mathrm{GP}},\Delta)$.  
%
%
In the asymptotic limit where $M_{\mathrm{GP}}L\gg 1$ it can be shown that 
(with $n\equiv |\vec{n}|$)
\bea
\label{eq:JFV_asymp}
 J_{\mathrm{FV}}(M_{\mathrm{GP}},\Delta) &=& 
\sum_{\vec{n}\not=\vec{0}} \left ( \frac{1}{8\pi n L}\right )
{\mathrm{e}}^{- n M_{\mathrm{GP}} L} 
{\mathcal{A}} ,
\eea
where
\bea
 {\mathcal{A}} &=&\exponential^{(z^{2})} \big [ 
1 - {\mathrm{Erf}}(z)\mbox{ }\big ]
+\left (\frac{1}{n M_{\mathrm{GP}} L} \right ) \bigg [
 \frac{1}{\sqrt{\pi}} \left ( \frac{z}{4} - 
\frac{z^{3}}{2}\right )
 + \frac{z^{4}}{2}\exponential^{(z^{2})} 
 \big [ 1 - {\mathrm{Erf}}(z)\mbox{ }\big ]
\bigg ]\nonumber\\
\label{eq:JFV_asymptotic}
& &
-\left (\frac{1}{n M_{\mathrm{GP}} L} \right )^{2}\bigg [
\frac{1}{\sqrt{\pi}}\left ( \frac{9z}{64} - 
\frac{5z^{3}}{32}
  +\frac{7z^{5}}{16} + \frac{z^{7}}{8} \right )
-\left ( \frac{z^{6}}{2} + \frac{z^{8}}{8}\right )
\exponential^{(z^{2})} \big [ 1 - {\mathrm{Erf}}(z)\mbox{ }\big ]
\bigg ]
+\op\left (\left [ \frac{1}{n M_{\mathrm{GP}} L}\right ]^{3}\right ) ,
\eea
%
with
\beq
z \equiv \left (\frac{\Delta}{M_{\mathrm{GP}}}\right ) 
\sqrt{\frac{n M_{\mathrm{GP}}L}{2}} .
\eeq
The quantity ${\mathcal{A}}$ is the alteration of finite volume
effects due to the presence of a non-zero $\Delta$.  
It multiplies the factor
exp($-n M_{\mathrm{GP}} L$), which results from the propagation
of the Goldstone particle to the boundary.
It is possible to analytically compute the higher order corrections 
of ${\mathcal{A}}$ in powers of 
$1/(n M_{\mathrm{GP}} L)$.  This way, one can achieve any 
desired numerical precision.  
Here it is clear that this alteration of volume effects is
controlled by the ratio in Eq.~(\ref{eq:mass_ratio}).   

%
%

\begin{figure}
\includegraphics[width=8.5cm]{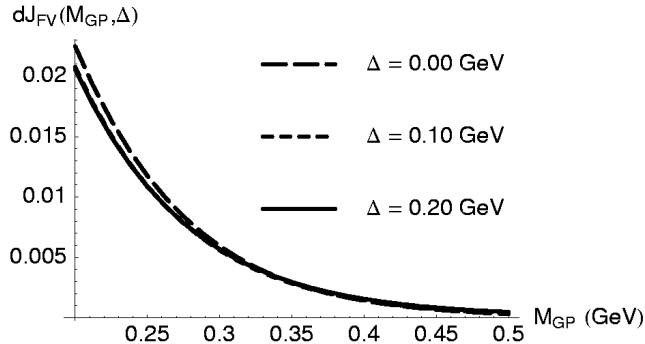}
\caption{\label{fig:dJFV}$dJ_{\mathrm{FV}}(M_{\mathrm{GP}},
\Delta)$, defined in Eq.~(\ref{eq:dJFV}), 
plotted against $M_{\mathrm{GP}}$, with three different choices of $\Delta$.
This function indicates the deviation
(in percent) of the asymptotic form of $J_{\mathrm{FV}}$ from the definition
in Eq.~(\ref{eq:JFV}). 
The size of the volume in this plot is 
$L=$2.5 fm.  The Goldstone mass $M_{\mathrm{GP}} = 0.197$ GeV corresponds
to $M_{\mathrm{GP}} L = 2.5$, and $M_{\mathrm{GP}} = 0.32$ GeV corresponds
to $M_{\mathrm{GP}} L = 4$.  
The curve for $\Delta=0.1$ GeV is hidden by that for $\Delta=0.2$ GeV.}
\end{figure}

\begin{figure}
\includegraphics[width=8.5cm]{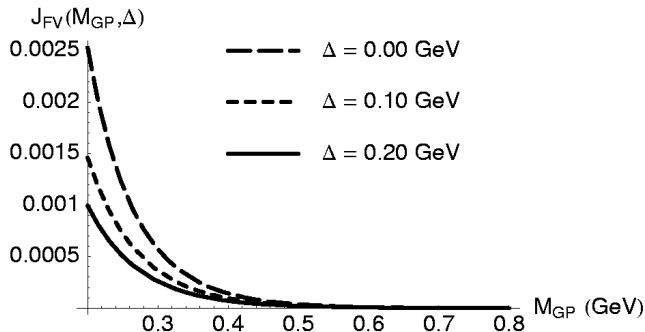}
\caption{\label{fig:JFV}$J_{\mathrm{FV}}(M_{\mathrm{GP}},
\Delta)$ 
plotted as 
a function of $M_{\mathrm{GP}}$, with three different choices of $\Delta$
corresponding to the three curves. The size of the volume in this plot is 
$L=$2.5 fm. The Goldstone mass $M_{\mathrm{GP}} = 0.197$ GeV corresponds
to $M_{\mathrm{GP}} L = 2.5$, and $M_{\mathrm{GP}} = 0.32$ GeV corresponds
to $M_{\mathrm{GP}} L = 4$.}
\end{figure}

Next, we consider different limits of ${\mathcal{A}}$ at  
fixed $M_{\mathrm{GP}}$ and $L$.  When
$\Delta=0$, clearly ${\mathcal{A}}=1$.  If
$\Delta$ is very small compared to $M_{\mathrm{GP}}$, such that 
$z\ll 1$, ${\mathcal{A}}$ is dominated by
the $(1/M_{\mathrm{GP}}L)^{0}$ term, {\it i.e.},
\beq
 {\mathcal{A}} \approx \exponential^{(z^{2})}
 \left [ 1-{\mathrm{Erf}}(z)\right ].
\eeq
Since Erf$(z)$ grows much faster than
exp($z^{2}$) in this regime, ${\mathcal{A}}$ will decrease as $\Delta$
increases.  When $\Delta$ is of $\op(M_{\mathrm{GP}})$ or 
larger, $z\gg 1$, and one can perform an asymptotic expansion of
the error function.  It can be shown that in this situation, 
\beq
 {\mathcal{A}} \sim \frac{1}{z} .
\eeq
That is, ${\mathcal{A}}$ also decreases as $\Delta$ increases.
We have also numerically checked that this is true when $z\approx 1$.
This means that the asymptotic formula in Eq.~(\ref{eq:JFV_asymp})
reproduces the physical picture outlined in the beginning of this section
for any $\Delta$. To demonstrate how fast the asymptotic form
in Eq.~(\ref{eq:JFV_asymptotic}) converges to 
Eq.~(\ref{eq:JFV}), we define 
\beq
\label{eq:dJFV}
 dJ_{\mathrm{FV}}(M_{\mathrm{GP}},\Delta) 
= \frac{J^{\mathrm{num}}_{\mathrm{FV}}(M_{\mathrm{GP}},\Delta)
 - J^{\mathrm{asymp}}_{\mathrm{FV}}(M_{\mathrm{GP}},\Delta)}
 {J^{\mathrm{num}}_{\mathrm{FV}}(M_{\mathrm{GP}},\Delta)} ,
\eeq
where $J^{\mathrm{num}}_{\mathrm{FV}}$ is the function $J_{\mathrm{FV}}$
evaluated numerically [Eq.~(\ref{eq:JFV})], and 
$J^{\mathrm{asymp}}_{\mathrm{FV}}$ is the asymptotic form in 
Eq.~(\ref{eq:JFV_asymptotic}).  In Fig.~\ref{fig:dJFV},
we plot $dJ_{\mathrm{FV}}$ as a function of $M_{\mathrm{GP}}$ with three
choices of $\Delta$.  It is clear from this plot that 
$J_{\mathrm{FV}}$ is
approximated well (to $\le 3\%$) by the asymptotic form
when $M_{\mathrm{GP}}L \ge 2.5$.  We use the asymptotic forms for
integrals of this type throughout this work.  Also, in this paper 
we only include  the terms with $|\vec{n}|=1,\sqrt{2}$, $\sqrt{3}$, 
$\sqrt{4}$ and $\sqrt{5}$ in the Poisson summation formula.  
We have confirmed that truncating the 
sum at $|\vec{n}|=\sqrt{5}$ is a very good approximation (to $\sim 3\%$)
when $M_{\mathrm{GP}} L \ge 2.5$.
The function $J_{\mathrm{FV}}(M_{\mathrm{GP}},\Delta)$ 
is plotted against $M_{\mathrm{GP}}$
in Fig.~\ref{fig:JFV}, with $L=2.5$ fm and three choices 
of $\Delta$.  It is clear from
this plot that $\Delta$ can significantly alter the 
finite volume effects in 
${\mathcal{J}}(M_{\mathrm{GP}},\Delta)$.
%

%
%

\begin{figure}
\includegraphics[width=8.5cm]{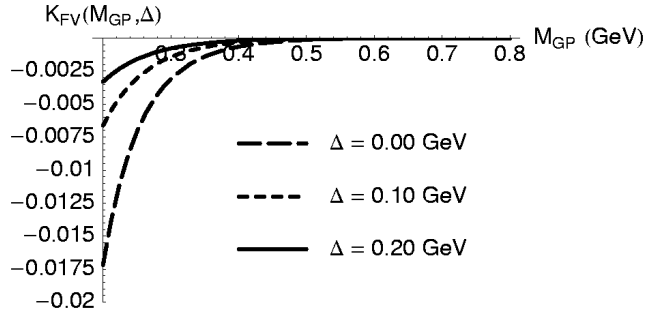}
\caption{\label{fig:KFV}$K_{\mathrm{FV}}(M_{\mathrm{GP}},
\Delta)$ 
plotted as 
a function of $M_{\mathrm{GP}}$, with three different 
choices of $\Delta$
corresponding to the three curves.  The size of the volume in this plot is 
$L=2.5$ fm.  The Goldstone mass $M_{\mathrm{GP}} = 0.197$ GeV corresponds
to $M_{\mathrm{GP}} L = 2.5$, and $M_{\mathrm{GP}} = 0.32$ GeV corresponds
to $M_{\mathrm{GP}} L = 4$ in this plot.}
\end{figure}

Another typical sum that appears in one-loop HM$\chi$PT in finite volume is
%
\beq
\label{eq:calK}
 {\mathcal{K}}(M_{\mathrm{GP}},\Delta) 
 = -i\left ( \frac{1}{L^{3}}\right )
 \sum_{\vec{k}} \int \frac{d k_{0}}{2\pi}
 \left (\frac{1}{k^{2}-M^{2}_{\mathrm{GP}} + i\epsilon} \right )
 \left (\frac{1}{v\cdot k - \Delta + i\epsilon} \right )^{2} .
\eeq
%
It is straightforward to show that
\bea
 {\mathcal{K}}(M_{\mathrm{GP}},\Delta) 
 = K(M_{\mathrm{GP}},\Delta) + 
 K_{\mathrm{FV}}(M_{\mathrm{GP}},\Delta) ,
\eea
where
\beq
 K(M_{\mathrm{GP}},\Delta) = 
 \frac{\partial J(M_{\mathrm{GP}},\Delta)}
 {\partial \Delta}
\eeq 
is the infinite volume limit of 
${\mathcal{K}}(M_{\mathrm{GP}},\Delta)$ and
\beq
\label{eq:KFV_asymptotic}
 K_{\mathrm{FV}}(M_{\mathrm{GP}},\Delta) = 
 \frac{\partial J_{\mathrm{FV}}(M_{\mathrm{GP}},\Delta)}
 {\partial \Delta}
\eeq
is the finite volume correction to $K(M_{\mathrm{GP}},\Delta)$.  
The function 
$K_{\mathrm{FV}}(M_{\mathrm{GP}},\Delta)$ is 
plotted against $M_{\mathrm{GP}}$
in Fig.~\ref{fig:KFV}, with $L=2.5$ fm and three choices of $\Delta$.  
As expected,
$|K_{\mathrm{FV}}(M_{\mathrm{GP}},\Delta)|$ also decreases when  
$\Delta$ increases
at fixed $M_{\mathrm{GP}}$ and $L$.

\section{Neutral $B$ mixing and heavy-light decay constants}
\label{sec:BBbar_mixing}
The study of neutral $B$ meson mixing allows the extraction of the 
magnitude of the CKM matrix element $V_{td}$, 
and hence the determination of one of the sides of the unitarity triangle.
The frequency of the $B_{d}{-}\bar{B}_{d}$ oscillations, which is given
by the mass difference, $\Delta m_{d}$, in this mixing system has been
well measured by various experimental collaborations 
\cite{Battaglia:2003in}.  It is also calculable in the standard model via
an operator product expansion in which the top quark and $W$ boson are 
integrated out.  Resumming the next-to-leading order (NLO) short-distance
QCD effects, one obtains
\bea
\label{eq:delta_md}
 \Delta m_{d} &=& \frac{G_{F}}{8 \pi^{2}} M^{2}_{W}
  |V_{td} V^{\ast}_{tb}|^{2} \eta_{B} S_{0}(x_{t}) C_{B}(\mu)
\nonumber\\
  & &\times
\frac{|\la\bar{B}_{d}|\op^{\Delta B = 2}_{d}(\mu)|B_{d}\ra|}{2 M_{B}}
 , 
\eea
where $\mu$ is the renormalisation scale, $x_{t}=m^{2}_{t}/M^{2}_{W}$, and 
$S_{0}(x_{t})\approx 0.784 x_{t}^{0.76}$ (to better than $1\%$) 
is the relevant Inami-Lim function \cite{Inami:1981fz}.  The 
coefficients $\eta_{B}=0.55$ and
$C_{B}(\mu)$ are from short-distance QCD effects
\cite{Buras:1990fn,Buchalla:1996vs}. The
matrix element of the four-quark operator 
\beq
\label{eq:4q_op_bd}
 \op^{\Delta B=2}_{d} = [\bar{b}\gamma^{\mu}(1 - \gamma_{5})d]
                [\bar{b}\gamma_{\mu}(1 - \gamma_{5})d]
\eeq
between $B_{d}$ and $\bar{B}_{d}$ states contains all the long-distance
QCD effects in Eq.~(\ref{eq:delta_md}), and has to be calculated 
non-perturbatively.   Since $|V_{tb}|=1$ to good
accuracy and $\Delta m_{d}$ has been well measured, a 
high-precision calculation of 
$\la\bar{B}_{d}|\op^{\Delta B = 2}_{d}(\mu)|B_{d}\ra$ enables 
a clean determination of $|V_{td}|$.

%
%

The frequency of the rapid $B_{s}{-}\bar{B}_{s}$ oscillations can be
precisely measured at the Tevatron and LHC~\cite{Battaglia:2003in}. 
Therefore an alternative approach
is to consider the ratio
\beq
\label{eq:delta_ms_over_delta_md}
 \frac{\Delta m_{s}}{\Delta m_{d}} =
 \left | \frac{V_{ts}}{V_{td}} \right |^{2} 
 \left ( \frac{M_{B_{d}}}{M_{B_{s}}}\right )
 \left | \frac{\la\bar{B}_{s}|\op^{\Delta B = 2}_{s}|B_{s}\ra}
  {\la\bar{B}_{d}|\op^{\Delta B = 2}_{d}|B_{d}\ra}
 \right | ,
\eeq
in which many theoretical uncertainties cancel.
Here $\Delta m_{s}$ is the mass difference in the $B_{s}{-}\bar{B}_{s}$
system and $\op^{\Delta B=2}_{s} = [\bar{b}\gamma^{\mu}
(1 - \gamma_{5})s][\bar{b}\gamma_{\mu}(1 - \gamma_{5})s]$.  The unitarity
of the CKM matrix implies $|V_{ts}|\approx |V_{cb}|$ to a few percent, and
$|V_{cb}|$ can be precisely extracted by analysing 
semileptonic $B$ decays \cite{Battaglia:2003in}.  Therefore a clean
measurement of $\Delta m_{s}/\Delta m_{d}$ will yield an accurate
determination of $|V_{td}|$.

%
%

The matrix elements in Eq.~(\ref{eq:delta_ms_over_delta_md}) are 
conventionally parameterised as
\beq
\label{eq:me_to_BfB}
\la\bar{B}_{q}|\op^{\Delta S = 2}_{q}|B_{q}\ra =
 \frac{8}{3} M^{2}_{B_{q}} f^{2}_{B_{q}} B_{B_{q}}(\mu) ,
\eeq
where the parameter $B_{B_{q}}$ measures the deviation from the 
vacuum-saturation approximation of the matrix element, and
$q=d$ or $s$.  The decay constant $f_{B_{q}}$ is defined by
\beq
 \la 0 |\bar{b}\gamma_{\mu}\gamma_{5}q|B_{q}(\vec{p})\ra =
 i p_{\mu} f_{B_{q}} .
\eeq
%

%
%

Lattice QCD provides a reliable tool for calculating these 
non-perturbative QCD quantities from first principles\footnote{Some
selected reviews in the long history of lattice calculations for the $B$ 
mixing system can be found in Refs.~\cite{Flynn:1998ca, 
Sachrajda:2000ci, Flynn:2000hx, Bernard:2000ki, Ryan:2001ej, Yamada:2002wh, 
Lellouch:2002nj, Becirevic:2003hf, Kronfeld:2003sd}.}.  
Since $\Delta m_{s}/\Delta m_{d}$ will be measured to very good accuracy,
it is important to have clean theoretical calculations for 
[the $SU(3)$ breaking ratios of] the matrix 
elements, decay constants and $B$ parameters involved. 
Current
lattice calculations have to be combined with effective theories in 
order to obtain these matrix elements at the physical quark masses.  
This procedure can introduce significant systematic errors and 
dominate the uncertainties in the $SU(3)$ breaking ratio
\cite{Kronfeld:2002ab,Becirevic:2002mh}
\beq
\label{eq:xi_def}
 \xi = \frac{f_{B_{s}} \sqrt{B_{B_{s}}}}{f_{B}\sqrt{B_{B}}} ,
\eeq
which is the key theoretical input for future high-precision determination
of $|V_{td}|$ via the study of neutral $B$ mixing\footnote{Notice that
the symbol $\xi$ as defined in Eq.~(\ref{eq:xi_def}) is in 
the traditional notation in $B$ physics, and has nothing to do with 
the Goldstone field $\xi$ introduced in Section \ref{sec:HMChPT}.}.
However, the use of effective theory also offers a framework 
for studying finite volume effects in lattice 
calculations~\cite{Bernard:1996ez, Golterman:1997wb, Sharpe:1992ft,
Golterman:1998st, Golterman:1998af, Lin:2002nq, Lin:2002aj, Lin:2003tn, 
Becirevic:2003wk, Detmold:2004qn, Beane:2003da, Beane:2003yx, 
Colangelo:2003hf, AliKhan:2003cu, beane:to_appear}.  
We will
demonstrate in this section that finite volume effects might turn out to
exceed the current quoted systematic errors for quantities such as $\xi$.

\subsection{The one-loop calculation in finite volume}
\label{sec:one_loop_calculation}
In this subsection, we discuss one-loop calculations for the $B$ 
parameters and heavy-light decay constants mentioned above in finite
volume HM$\chi$PT including the appropriate mass shifts to the first
non-trivial order of the chiral and $1/M_{P}$ expansion.  The inclusion
of other first-order corrections in these quantities
is straightforward.  It simply introduces additional
low-energy constants (LECs)
which account for short-distance physics and do not give rise to finite volume
effects at this order, so we will not discuss this issue here.
We have performed the calculation for full QCD, QQCD and PQQCD
with the mass shifts given between
Eqs.~(\ref{eq:shift_start}) and (\ref{eq:shift_end}).

%
%

For the purpose of this work,
the axial current $\bar{b}\gamma_{\mu}\gamma_{5}q_{a}$ is 
\beq
\label{eq:axial_current}
 A_{\mu} = \left ( \frac{\kappa}{2}\right ) {\mathrm{tr}}_{\mathrm{D}}
 \left [ \gamma_{\mu}\gamma_{5} H^{(Q)}_{b}\xi^{\dagger b}_{a}\right ],
\eeq
and the four-quark operator $\op^{\Delta P_{a}=2}$ (when $P_{a}=B_{d,s}$, 
$\op^{\Delta P_{a}=2}$ becomes $\op^{\Delta B=2}_{d,s}$) is
\bea
\label{eq:Oaa}
 O^{aa} &=& 4\beta \left [ 
 \left ( \xi P^{\ast(Q)\dagger}_{\mu}\right )^{a}
 \left ( \xi P^{\ast(\bar{Q})\mu}\right )^{a}
\right. \nonumber\\
 &&\left. +
 \left ( \xi P^{(Q)\dagger}\right )^{a}
 \left ( \xi P^{(\bar{Q})}\right )^{a}
\right ] 
\eea
in HM$\chi$PT \cite{Grinstein:1992qt}, where $\kappa$ and $\beta$ are 
the low-energy constants which have to be determined
from experiments or lattice calculations.  
Notice that the index
$a$ in Eq.~(\ref{eq:Oaa}) is not summed over.  
Again, the inclusion of the chiral and $1/M_{P}$ corrections in these
operators simply introduces additional LECs and we do not investigate
this aspect here.
We assume that $\kappa$ and $\beta$ are the same in full, quenched
and partially quenched QCD.  Also, $A_{\mu}$ and $O^{aa}$ can couple
to the $\eta^{\prime}$ in QQCD, but the couplings are $1/N_{c}$
suppressed \cite{Sharpe:1996qp}, and we neglect them.

%
%
%
%
\begin{figure}
\includegraphics[width=8.5cm]{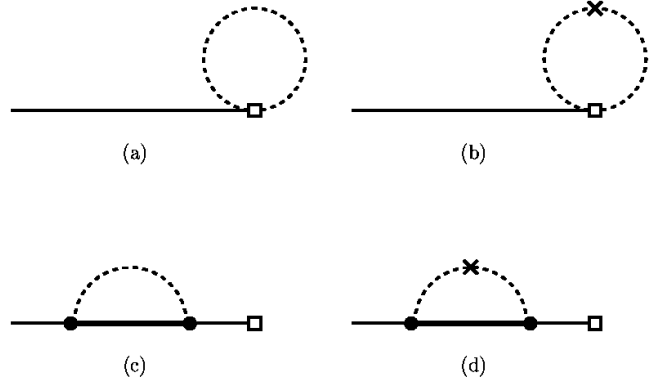}
\caption{\label{fig:fB_diagrams}Diagrams contributing to the one-loop
calculation of decay constants.  The thin (thick) solid 
lines are 
the heavy-light pseudoscalar (vector) mesons.  The dashed 
lines are Goldstone particles, and the crosses are the ``double poles''
which appear in (P)Q$\chi$PT.  The open squares are the 
operators defined in Eq.~(\ref{eq:axial_current}) and the dots are
vertices from the HM$\chi$PT Lagrangian.  Diagrams (c) and (d) are 
for wavefunction renormalisation.}
\vspace{0.1cm}
\end{figure}
%
\begin{figure}
\includegraphics[width=8.5cm]{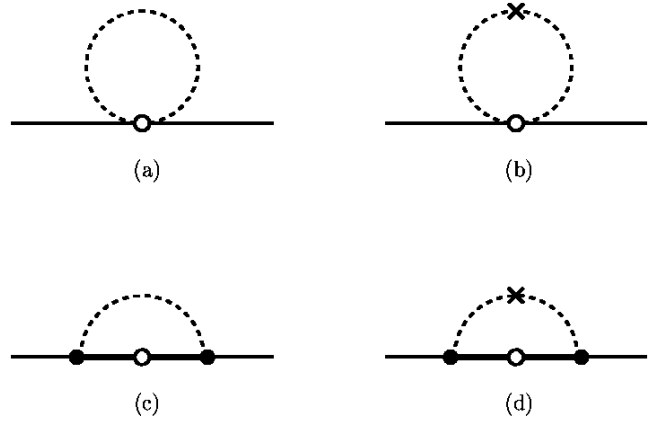}
\caption{\label{fig:BB_diagrams}Diagrams contributing to the one-loop
calculation of the $B$ parameters.   The open circles are the 
operators defined in Eq.~(\ref{eq:Oaa}).}
\vspace{0.1cm}
\end{figure}
The diagrams contributing to $f_{P_{(s)}}$ and $B_{P_{(s)}}$ are 
presented in Figs. \ref{fig:fB_diagrams} and \ref{fig:BB_diagrams}
respectively.  Only diagrams (c) and (d) depend on the heavy meson
mass shifts.  Therefore it is the relative weight between these diagrams
and the ``tadpole'' diagrams (a) and (b) which determines the dependence
on the heavy meson mass in finite volume effects.  

%
%

Although this is the first one-loop calculation for these decay constants and 
$B$ parameters in finite volume,
some results in the infinite volume limit
already exist in the literature: $f_{P_{(s)}}$ have been calculated at the
lowest order in full 
\cite{Grinstein:1992qt}, quenched \cite{Sharpe:1996qp,Booth:1995hx}
and partially quenched \cite{Sharpe:1996qp} QCD, and up to first-order
corrections in the chiral and
$1/M_{P}$ expansions in full \cite{Boyd:1995pa} and 
quenched QCD \cite{Booth:1994rr}.  The $B$ parameters have been
calculated only at 
lowest order \cite{Grinstein:1992qt, Sharpe:1996qp}.  Our results, 
as presented in Appendix \ref{sec:results}, agree 
with all these previous calculations in the appropriate limits.

\subsection{Phenomenological impact}
\label{sec:phenomenology}
We have used the one-loop results in Appendix \ref{sec:results}
to investigate the impact of finite volume effects on $\xi$.  
In this work, we only intend to estimate
the possible size of errors in this quantity, and will
leave the actual comparison with lattice data to a future publication.
Notice that the one-loop calculation is only valid for
$M_{\pi} \ll \Lambda_{\chi}$.  Nevertheless, we present our 
results for $M_{\pi}$ up to $\sim 800\mev$, where, in principle, 
higher-order chiral corrections should be included.  However, 
finite volume effects are exponentially suppressed at such large
pion masses.

%
%

Following the usual procedure in lattice calculations for $\xi$, 
we study two $SU(3)$ breaking ratios
\beq
\label{eq:xi_f}
 \xi_{f} = \frac{f_{B_{s}}}{f_{B}} 
\eeq
and
\beq
\label{eq:xi_B}
 \xi_{B} = \frac{B_{B_{s}}}{B_{B}} ,
\eeq
in terms of which,  
\beq
 \xi = \xi_{f}\sqrt{\xi_{B}} .
\eeq
Furthermore, we define 
\beq
 (\xi_{f})_{\mathrm{FV}}\mbox{ }{\mathrm{and}}\mbox{ }
 (\xi_{B})_{\mathrm{FV}}
\eeq
to be the contributions from finite volume effects, {\it i.e.}, those from
the volume-dependent part in the one-loop results presented in Subsection
\ref{sec:one_loop_calculation}.  To be more precise,
we use the formulae collected in Appendix \ref{sec:results} to calculate
the volume corrections {\it with respect to the lowest-order values} of 
$f_{B_{s}}$ ($B_{B_{s}}$) and $f_{B}$ ($B_{B}$), then take the difference
between the results as an estimate of $(\xi_{f})_{\mathrm{FV}}$
[$(\xi_{B})_{\mathrm{FV}}$].  Since these $SU(3)$ ratios are not very
different from unity (at most $\sim 20\%$), this is a reasonable estimate
of these effects.  

%
%

Traditionally, many quenched lattice simulations of $B_{B_{(s)}}$ and
$f_{B_{(s)}}$ were performed using $L \sim 1.6$~fm.  Therefore
we present our estimate for finite volume effects 
in QQCD with this box size.  For comparison,
we adopt the same volume for full QCD.  
As for PQQCD, we work with $L = 2.5$~fm where most current
high-precision simulations are carried out~\cite{Davies:2003ik}.
Throughout this subsection, we ensure that the condition
\beq
 M_{\pi} L \ge 2.5 
\eeq
holds in all the plots presented here.

%
%

%
%
\begin{figure}
\includegraphics[width=8.5cm]{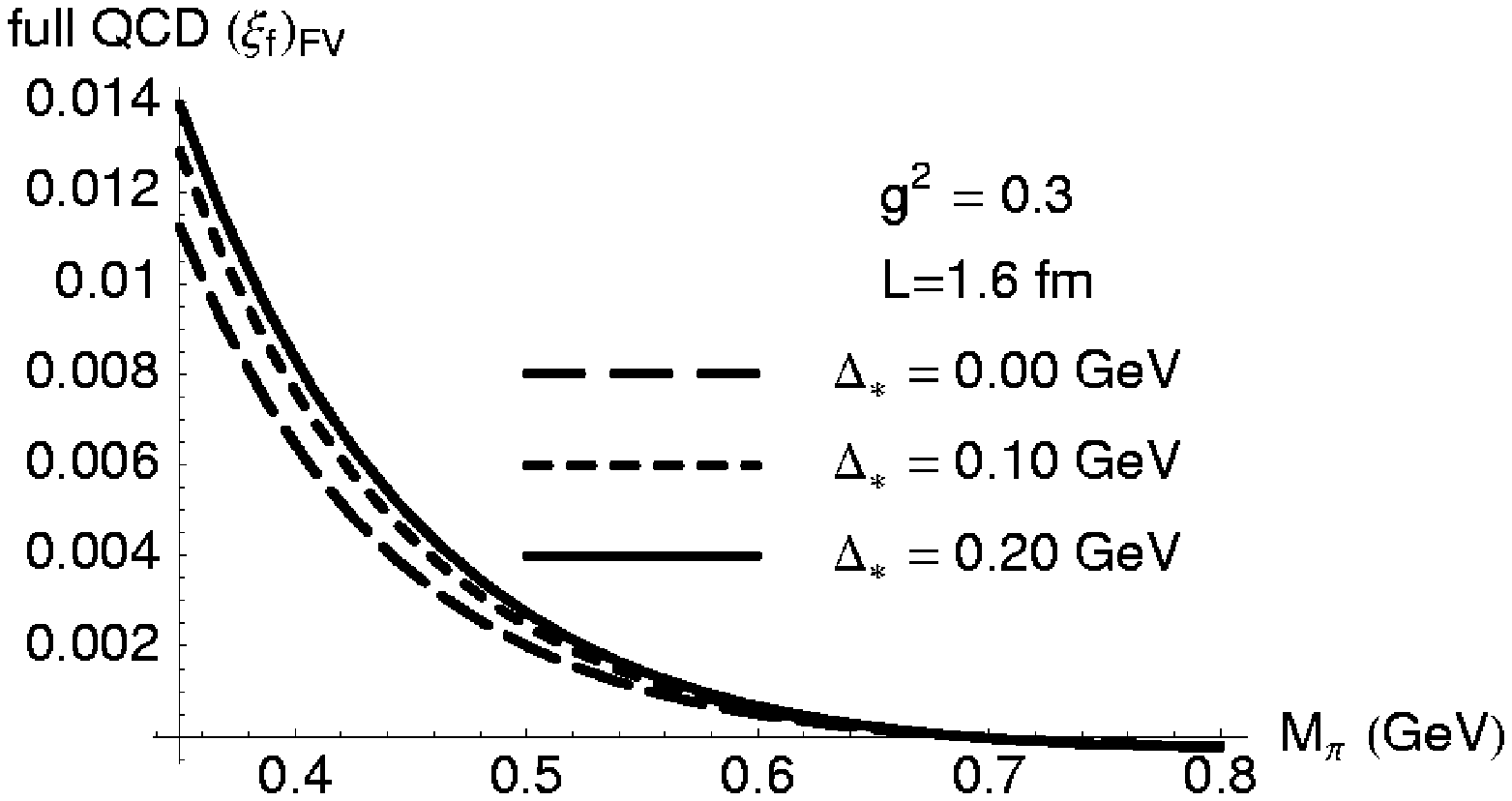}
\includegraphics[width=8.5cm]{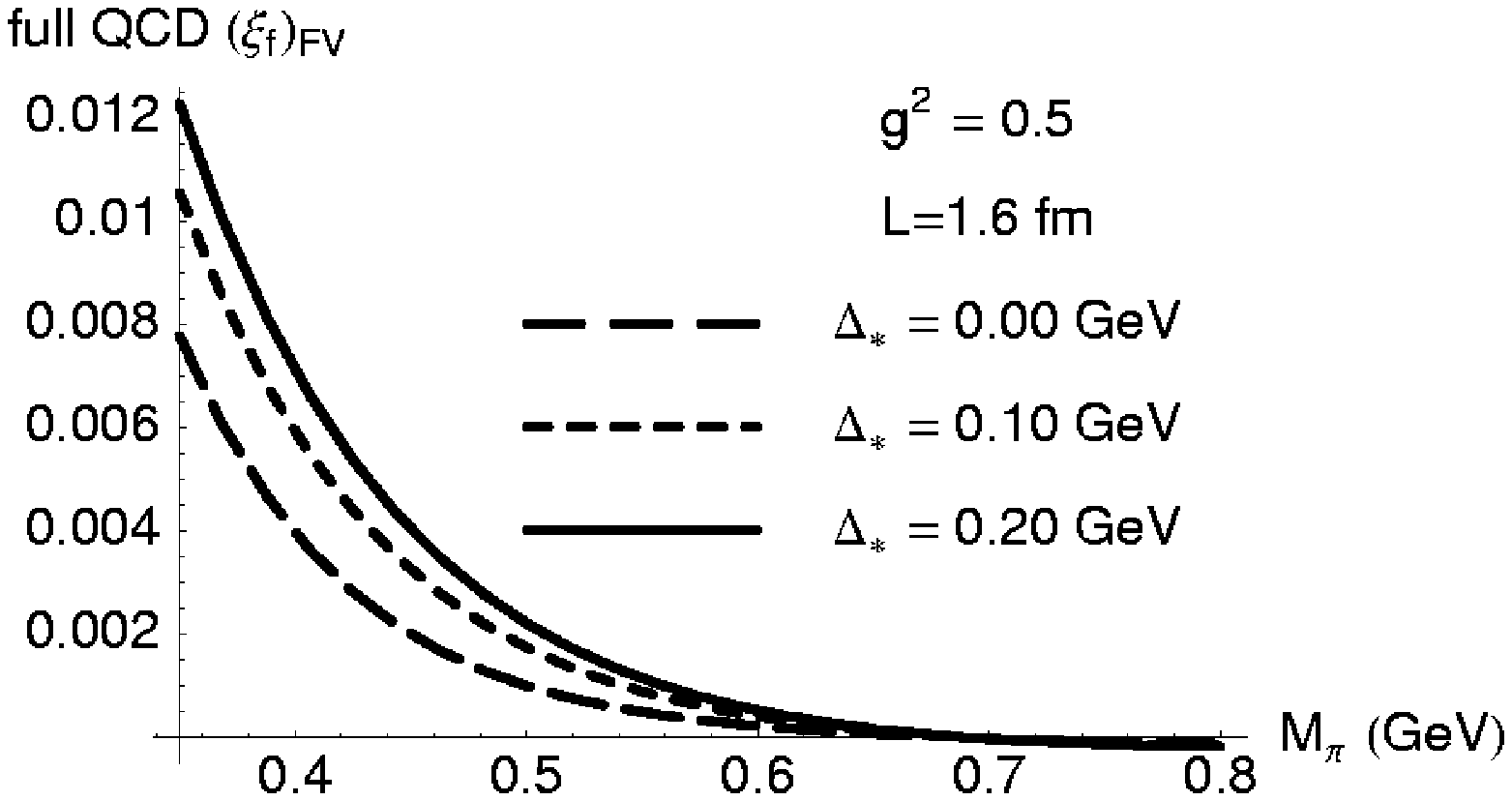}
\caption{\label{fig:fBs_over_fB}$(\xi_{f})_{\mathrm{FV}}$ in 
full QCD plotted against
$M_{\pi}$, with $L=1.6$ fm. 
The pion mass $M_{\pi} = 0.35$ GeV corresponds
to $M_{\pi} L = 2.8$, and $M_{\pi} = 0.5$ GeV corresponds
to $M_{\pi} L = 4$ in this plot.}
\end{figure}
\begin{figure}
\includegraphics[width=8.5cm]{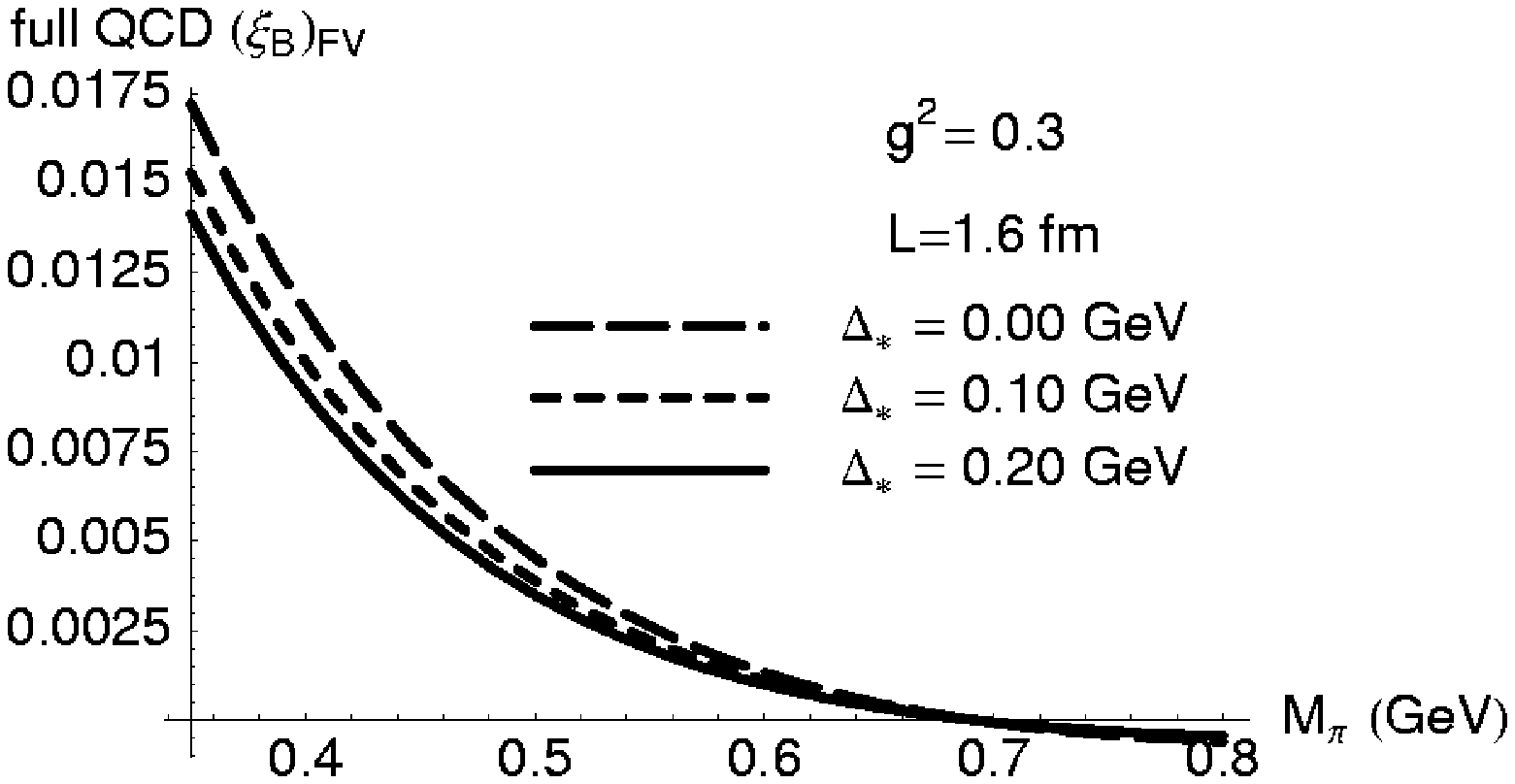}
\includegraphics[width=8.5cm]{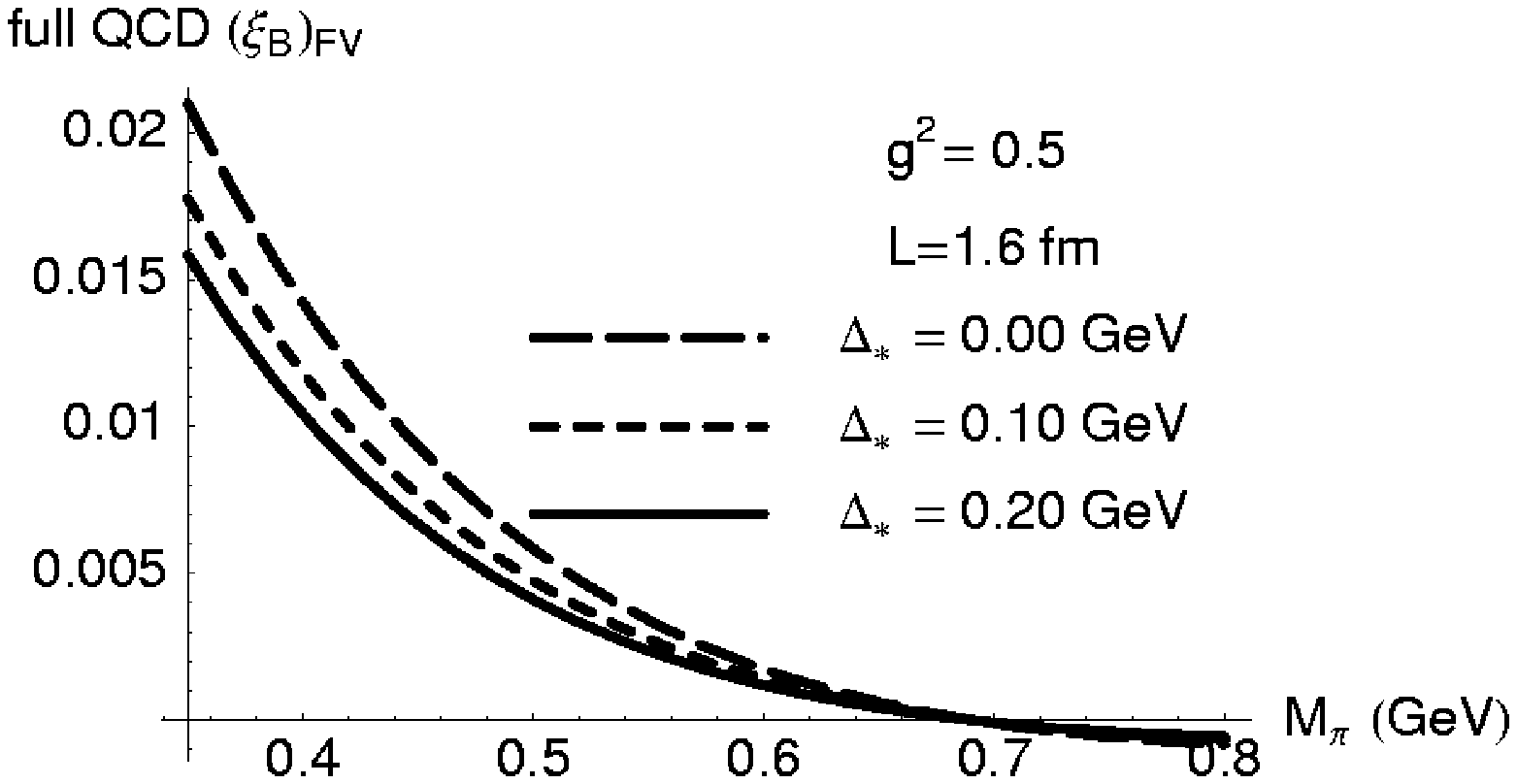}
\caption{\label{fig:BBs_over_BB}$(\xi_{B})_{\mathrm{FV}}$ in full QCD
plotted against
$M_{\pi}$, with $L=1.6$ fm. 
The pion mass  $M_{\pi} = 0.35$ GeV corresponds
to $M_{\pi} L = 2.8$, and $M_{\pi} = 0.5$ GeV corresponds
to $M_{\pi} L = 4$ in this plot.}
\end{figure}
%

We first discuss the procedure in full and quenched QCD.
When studying
the light quark mass dependence of
$(\xi_{f})_{\mathrm{FV}}$ and $(\xi_{B})_{\mathrm{FV}}$, we follow a 
strategy similar to that in Ref.~\cite{Kronfeld:2002ab}.  That is, we
use the Gell-Mann-Okubo formulae to express
$M_{K}$ and $M_{\eta}$ in terms of $M_{\pi}$ and $M_{33}$ ($M_{33}$,
defined in Eq.~(\ref{eq:m33}), is the mass of the fictitious
meson composed of $s$ and $\bar{s}$ quarks)
\beq
 M^{2}_{K} = \frac{M^{2}_{33} + M^{2}_{\pi}}{2} ,
\eeq
and
\beq
 M^{2}_{\eta} = \frac{2 M^{2}_{33} + M^{2}_{\pi}}{3} .
\eeq
We investigate the situation where a lattice calculation is performed at the
physical strange quark mass $(m_{s})_{\mathrm{phys}}$, 
but the up and down quark mass $m$ is varied.
By using $(M_{K})_{\mathrm{phys}}=0.498$ GeV and 
$(M_{\pi})_{\mathrm{phys}}=0.135$ GeV 
\cite{Hagiwara:2002fs}, we
fix,
\beq
 (M_{33})_{\mathrm{phys}} = 2 B_{0} (m_{s})_{\mathrm{phys}} = 0.691 
 \mbox{ }{\mathrm{GeV}}  ,
\eeq
as an input parameter in our analysis.  Notice that $(M_{33})_{\mathrm{phys}}$
is not the mass of a ``physical'' meson, and the subscript just means this
mass is 
estimated by using physical kaon and pion masses.
To the same order, we
can adopt Eq.~(\ref{eq:delta_s_lambda}) to write
\beq
 \delta_{s} = \lambda_{1} \left ( M^{2}_{33} - M^{2}_{\pi}\right ) ,
\eeq
and use $(M_{33})_{\mathrm{phys}}$, $(M_{\pi})_{\mathrm{phys}}$ and physical 
$M_{B_{s}}-M_{B}=0.091$ GeV \cite{Hagiwara:2002fs} to 
determine
\beq
\label{eq:lambda_value}
  \lambda_{1} = 0.1982 \gev^{-1} .
\eeq
This determines how $\delta_{s}$ varies with $M_{\pi}$.
We have also tried to use vanishing pion mass and $M_{D_{s}}-M_{D}=0.104$
GeV \cite{Hagiwara:2002fs} to fix $(M_{33})_{\mathrm{phys}}$ 
and $\lambda_{1}$, and the results
presented in this subsection are not sensitive to this
variation from the values quoted above.

%
%

%
%
\begin{figure}
\includegraphics[width=8.5cm]{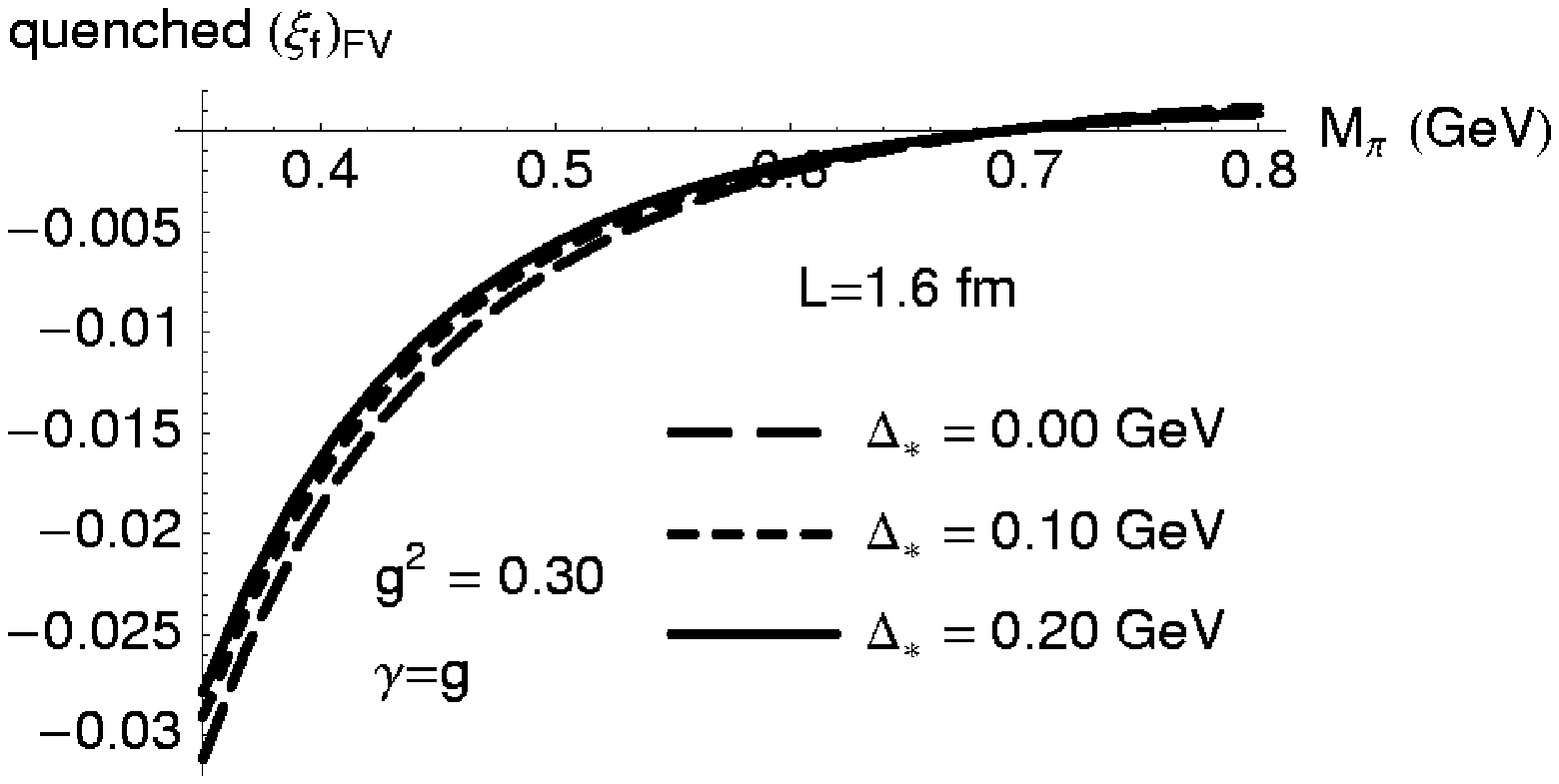}
\includegraphics[width=8.5cm]{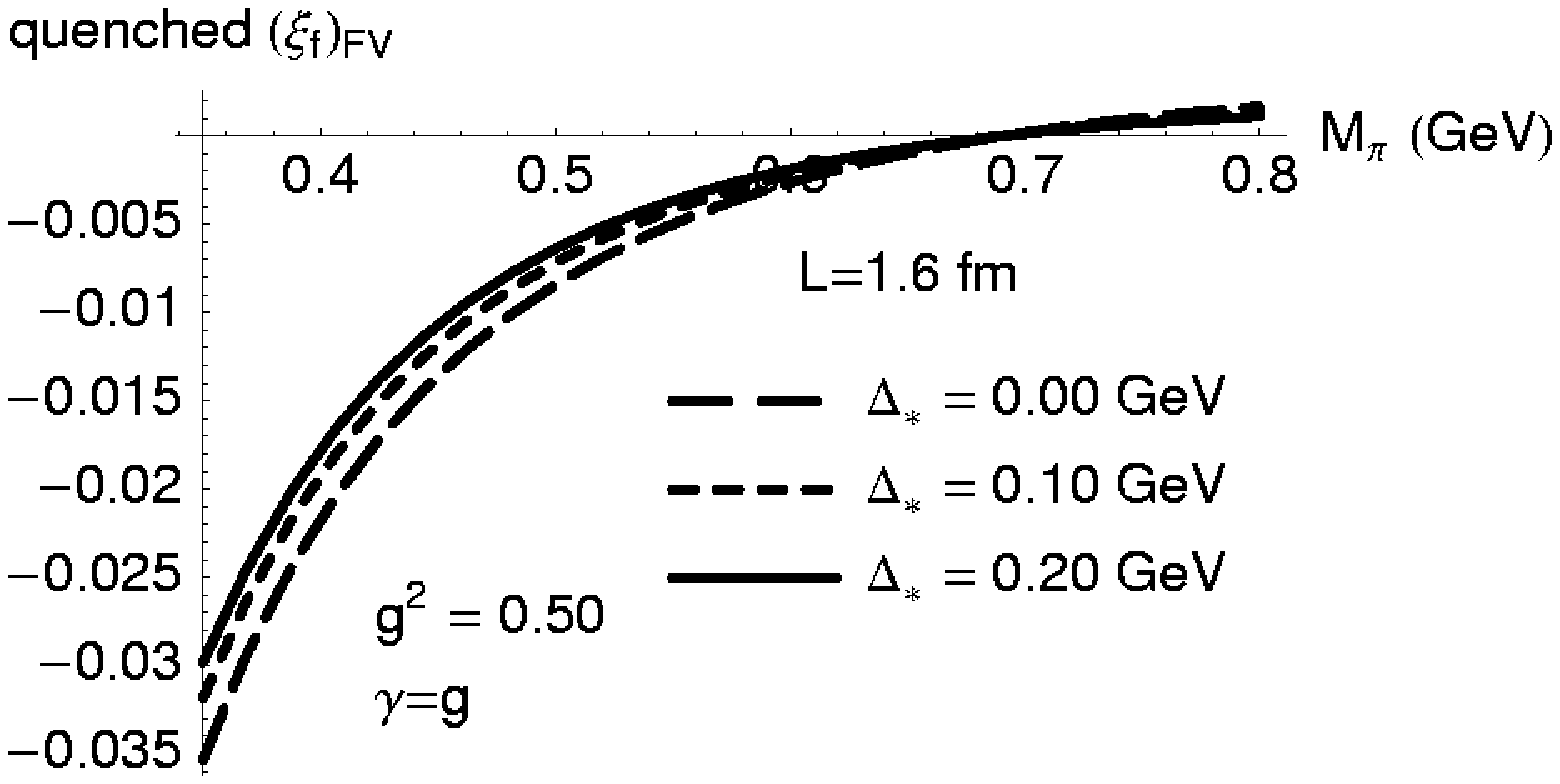}
\includegraphics[width=8.5cm]{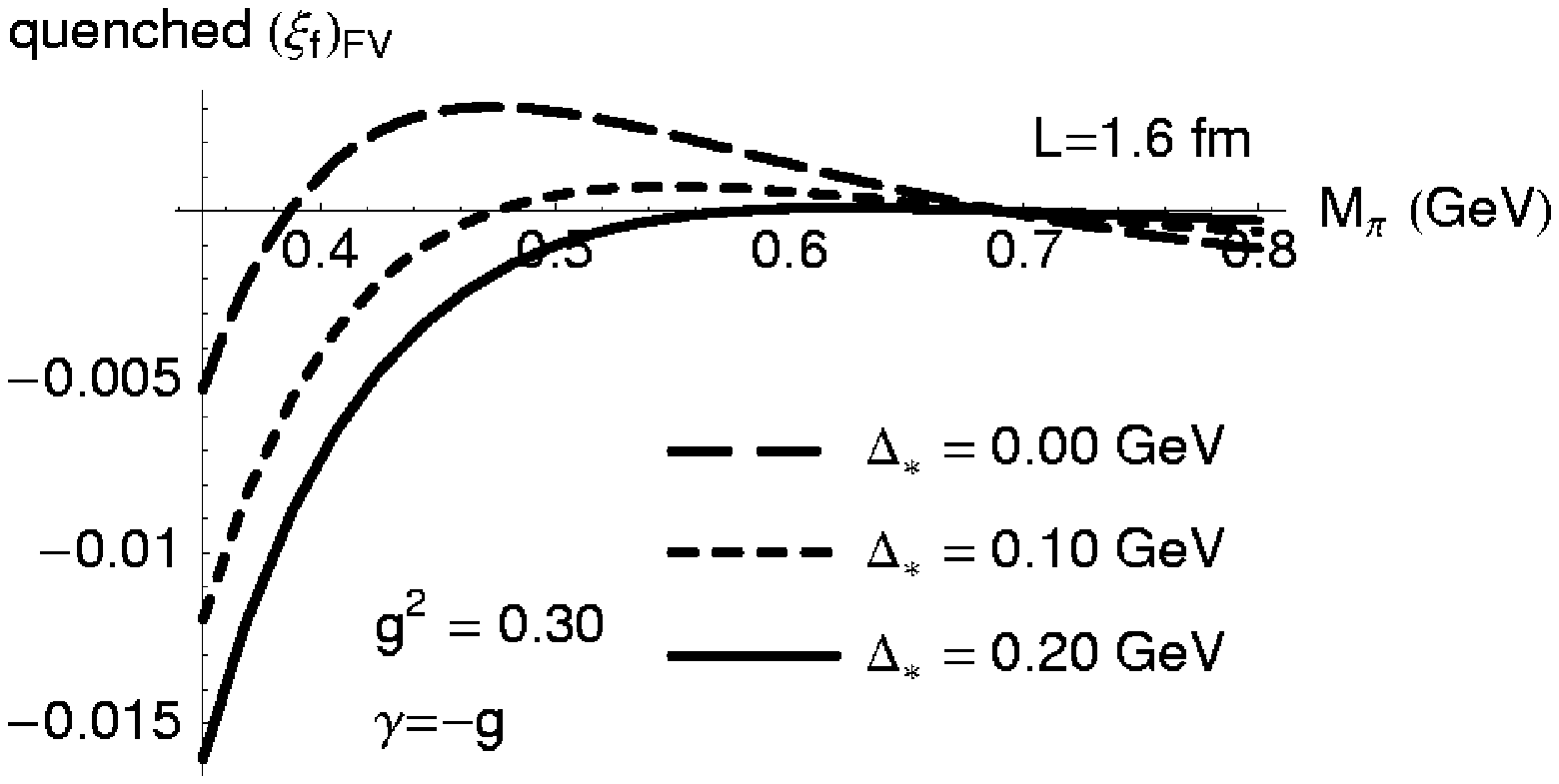}
\includegraphics[width=8.5cm]{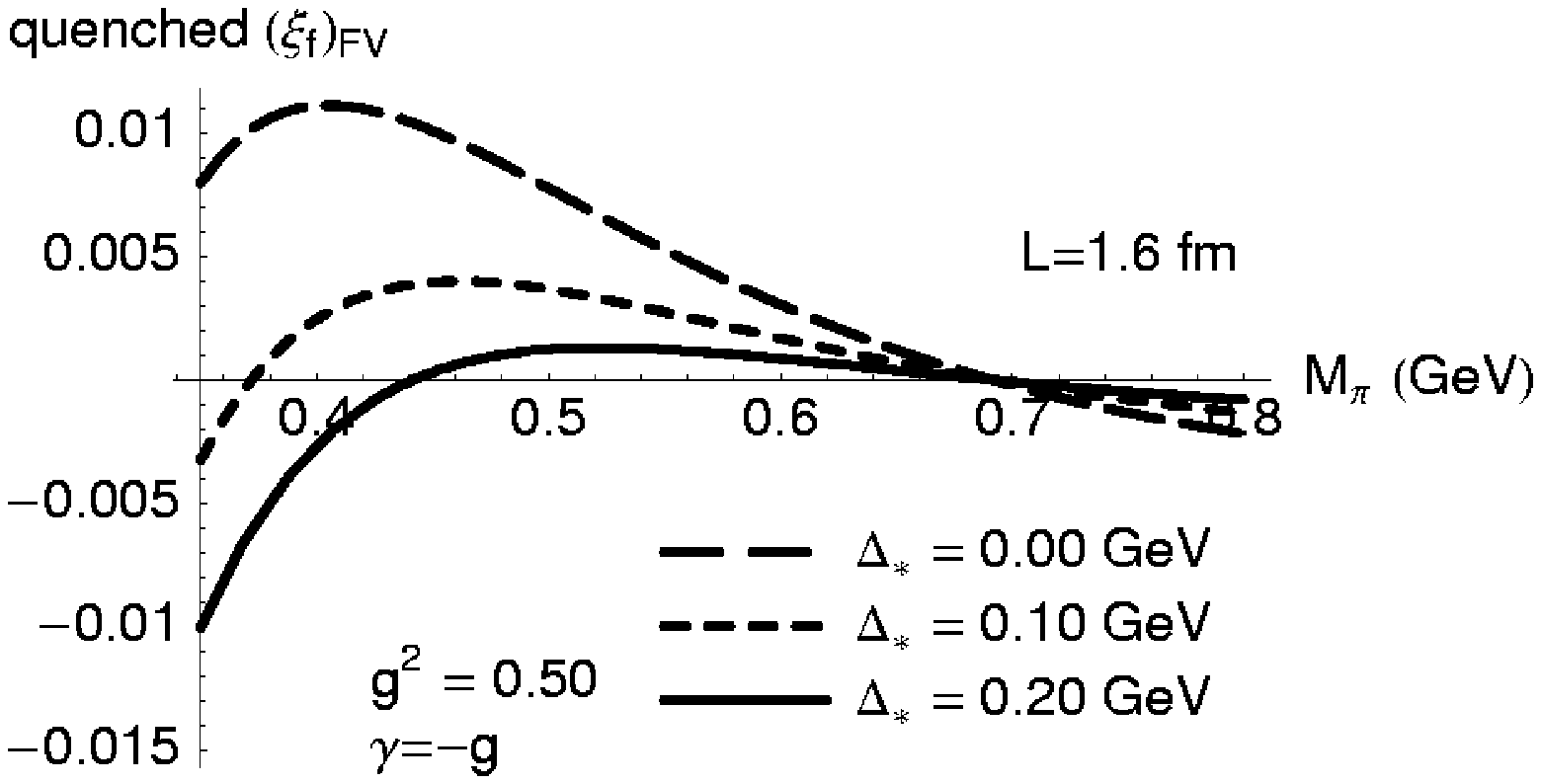}
\caption{\label{fig:Q_fBs_over_fB}$(\xi_{f})_{\mathrm{FV}}$ 
in QQCD plotted against
$M_{\pi}$, with $L=1.6$ fm and choices of the couplings 
$g$ and $\gamma$.  The pion mass $M_{\pi} = 0.35$ GeV corresponds
to $M_{\pi} L = 2.8$, and $M_{\pi} = 0.5$ GeV corresponds
to $M_{\pi} L = 4$ in this plot.
We set $\alpha=0$ and $M_{0}=700$ MeV.}
\end{figure}
\begin{figure}
\includegraphics[width=8.5cm]{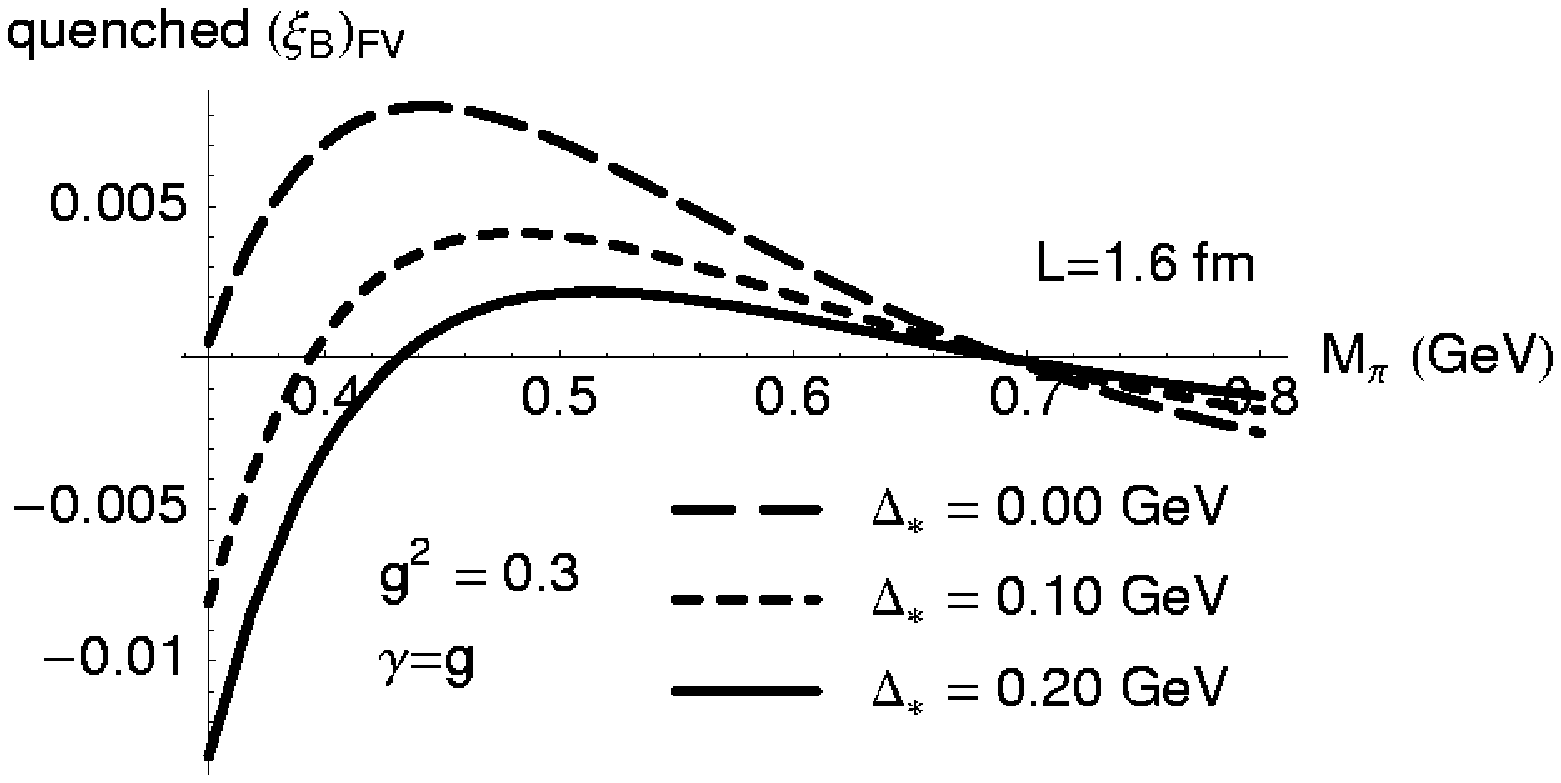}
\includegraphics[width=8.5cm]{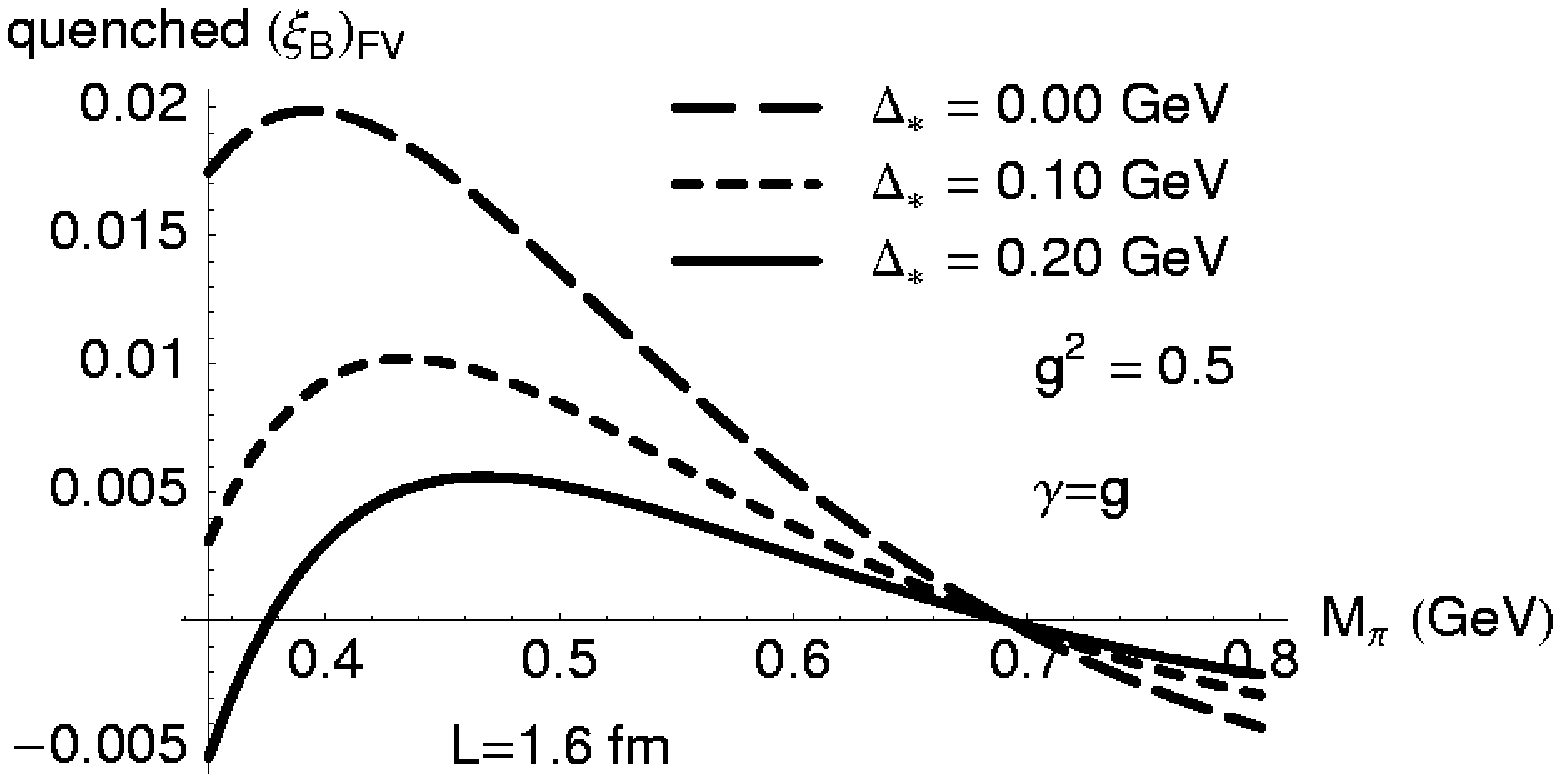}
\includegraphics[width=8.5cm]{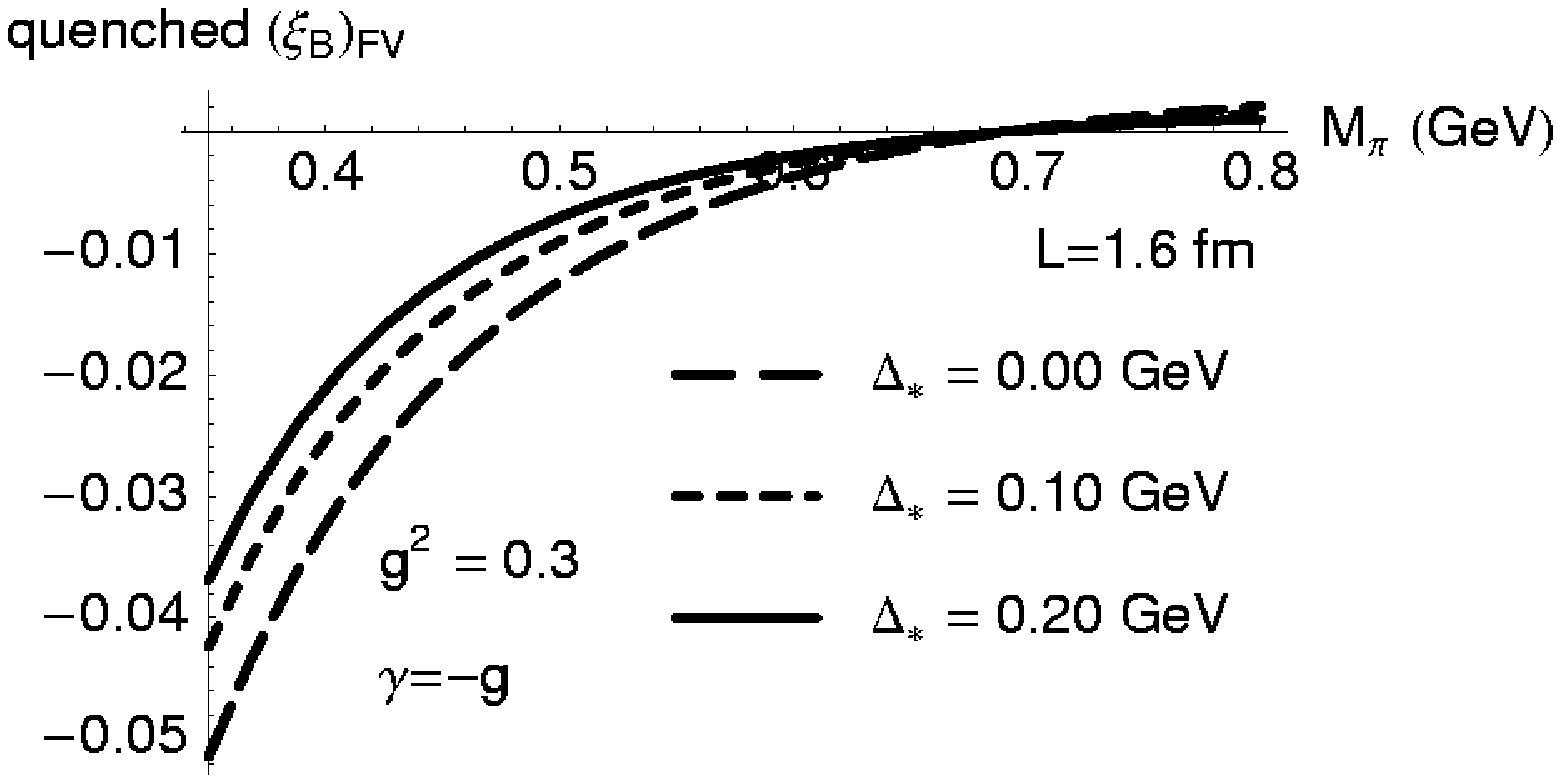}
\includegraphics[width=8.5cm]{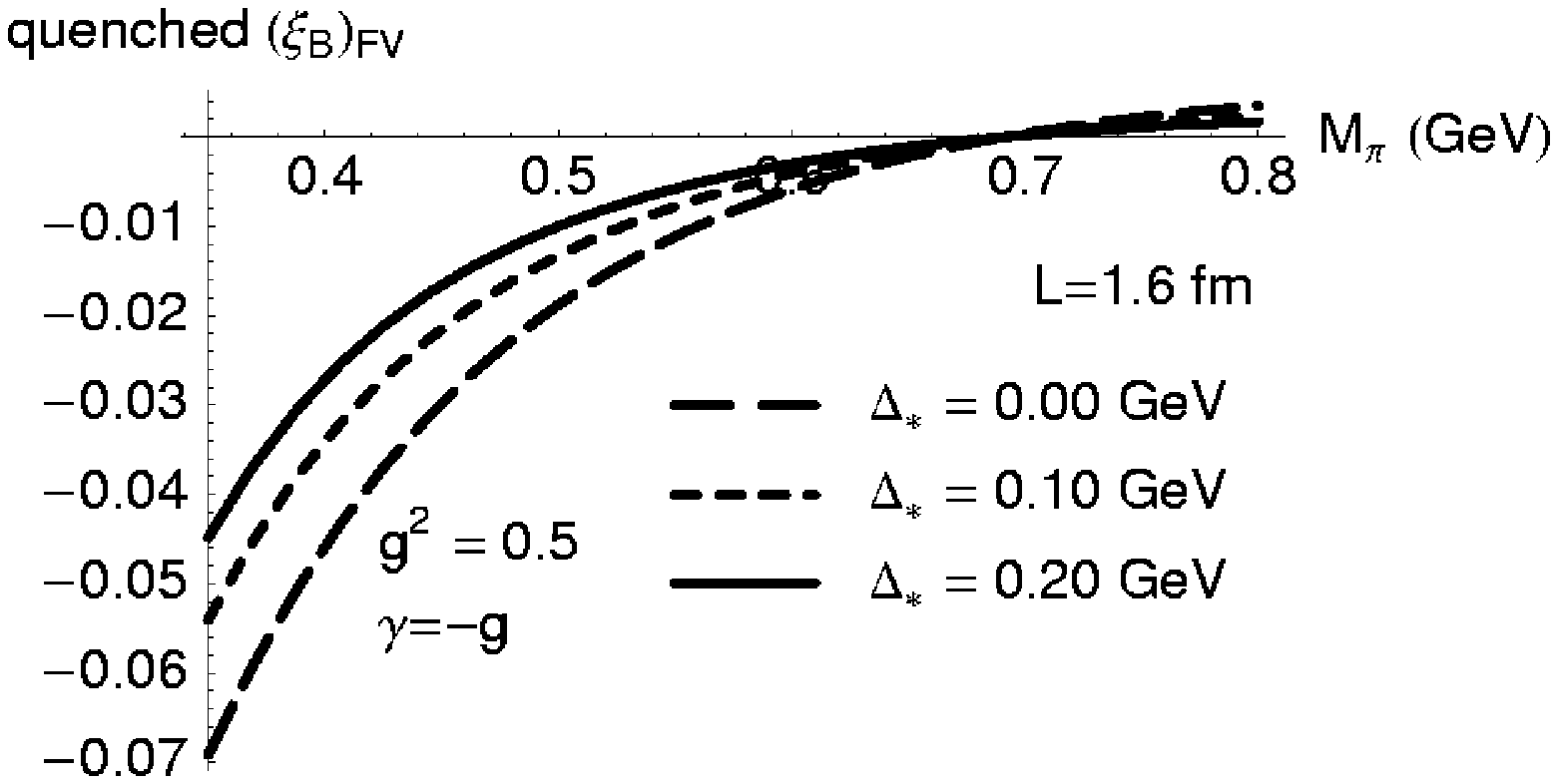}
\caption{\label{fig:Q_BBs_over_BB}$(\xi_{B})_{\mathrm{FV}}$ 
in QQCD plotted against
$M_{\pi}$, with $L=1.6$ fm and choices of the couplings 
$g$ and $\gamma$. The pion mass $M_{\pi} = 0.35$ GeV corresponds
to $M_{\pi} L = 2.8$, and $M_{\pi} = 0.5$ GeV corresponds
to $M_{\pi} L = 4$ in this plot.  
We set $\alpha=0$ and $M_{0}=700$ MeV.}
\end{figure}
%
%
The results for $(\xi_{f})_{\mathrm{FV}}$ and 
$(\xi_{B})_{\mathrm{FV}}$ for full QCD and QQCD from this analysis
are presented in Figs.~\ref{fig:fBs_over_fB}--\ref{fig:Q_BBs_over_BB}, 
with two different values 
for the coupling $g$ (and also $\gamma$ in QQCD).  
Here we stress again that the influence on finite volume effects from
the presence of $\Delta_{\ast}$ and $\delta_{s}$ depends on the size of 
these couplings, which are not well determined.  Inspired by the recent
CLEO measurement of $g$ in the charm system 
\cite{Ahmed:2001xc,Anastassov:2001cw}, and a recent
lattice calculation \cite{Abada:2003un}, we vary $g^{2}$ between 0.3
and 0.5. As for the coupling $\gamma$, which is a quenching artifact and 
has never been determined, we
vary its value between $g$ and $-g$.
It is clear from these plots that the finite volume
effects are generally small in full QCD ($\le 2\%$), 
but can be significant in QQCD ($\sim 3\%$ to $\sim 7\%$ for $\xi_{B}$) 
in the range of
$M_{\pi} L$ where lattice simulations are normally performed.  This is clearly
due to the enhanced long-distance effect arising from the ``double pole''
structure in (P)QQCD, as first pointed out in Ref.~\cite{Bernard:1996ez},
and manifests itself in various places, {\it e.g.}, nucleon-nucleon
potentials \cite{Beane:2002vq} and $\pi{-}\pi$ scattering
\cite{Bernard:1996ez, Golterman:1999hv, Lin:2003tn, Lin:2002aj}.

%
%

Although it has been well established that 
infinite volume chiral corrections are smaller in the $B$ parameters
than in the decay constants due to the coefficient in front of
$g^{2}$ in the one-loop results, it is clear from these plots that
finite volume effects are more salient in $\xi_{B}$ than in $\xi_{f}$.
All the quenched lattice calculations for $\xi_{B}$ have so far concluded that
this quantity 
is consistent with unity with typically $3\%$ error.
However, we find that the volume effects
are already at the level of $3{-}4\%$ when $M_{\pi}=0.45$ GeV in
a 1.6 fm box where
many quenched simulations were carried out.  
This error depends on both light and heavy quark masses in the simulation, 
hence is amplified after 
extrapolating the result to the physical quark masses.  
Also, the fact
that volume effects tend towards different directions in full QCD and QQCD
when $M_{\pi}$ becomes
smaller indicates
that quenching errors in these quantities can be larger than those
estimated in Ref.~\cite{Sharpe:1996qp}.
Since finite volume effects have not been included in the analysis of 
lattice calculations of $\xi_{B}$ hitherto, one should be cautious 
when using the existing quenched results for this quantity in 
any phenomenological work.

%
%

For the analysis in PQQCD, we assume that both the valence and sea
strange quark masses are
fixed at that of the physical strange quark.  
However, we vary the light sea quark mass $\tilde{m}$.  For this
purpose, we define
$M_{SS}$ to be the mass of the meson composed of two light sea
quarks.  Therefore,
\beq
 \frac{M^{2}_{SS}}{(M_{33})^{2}_{\mathrm{phys}}} = 
  \left (\frac{\tilde{m}}{\tilde{m}_{s}} 
  \right )_{\tilde{m}_{s}\mbox{ }=\mbox{ }{\mathrm{physical}}\mbox{ }m_{s}}.
\eeq
Also, we can express the mass shifts 
$\tilde{\delta}_{s}$ and 
$\delta_{\mathrm{sea}}$ in terms of meson masses:
\beq
 \tilde{\delta}_{s} = \lambda_{1} \left (M_{33}^{2}-M_{SS}^{2} \right )
\eeq
and
\beq
 \delta_{\mathrm{sea}} = \lambda_{1} \left (M^{2}_{SS}-M^{2}_{\pi} \right )
\eeq
by using Eqs.~(\ref{eq:tilde_delta_s_def}) and (\ref{eq:delta_sea_def})
with the same value of $\lambda_{1}$ as in Eq.~(\ref{eq:lambda_value}).

%
%

%
\begin{figure}
\includegraphics[width=8.5cm]{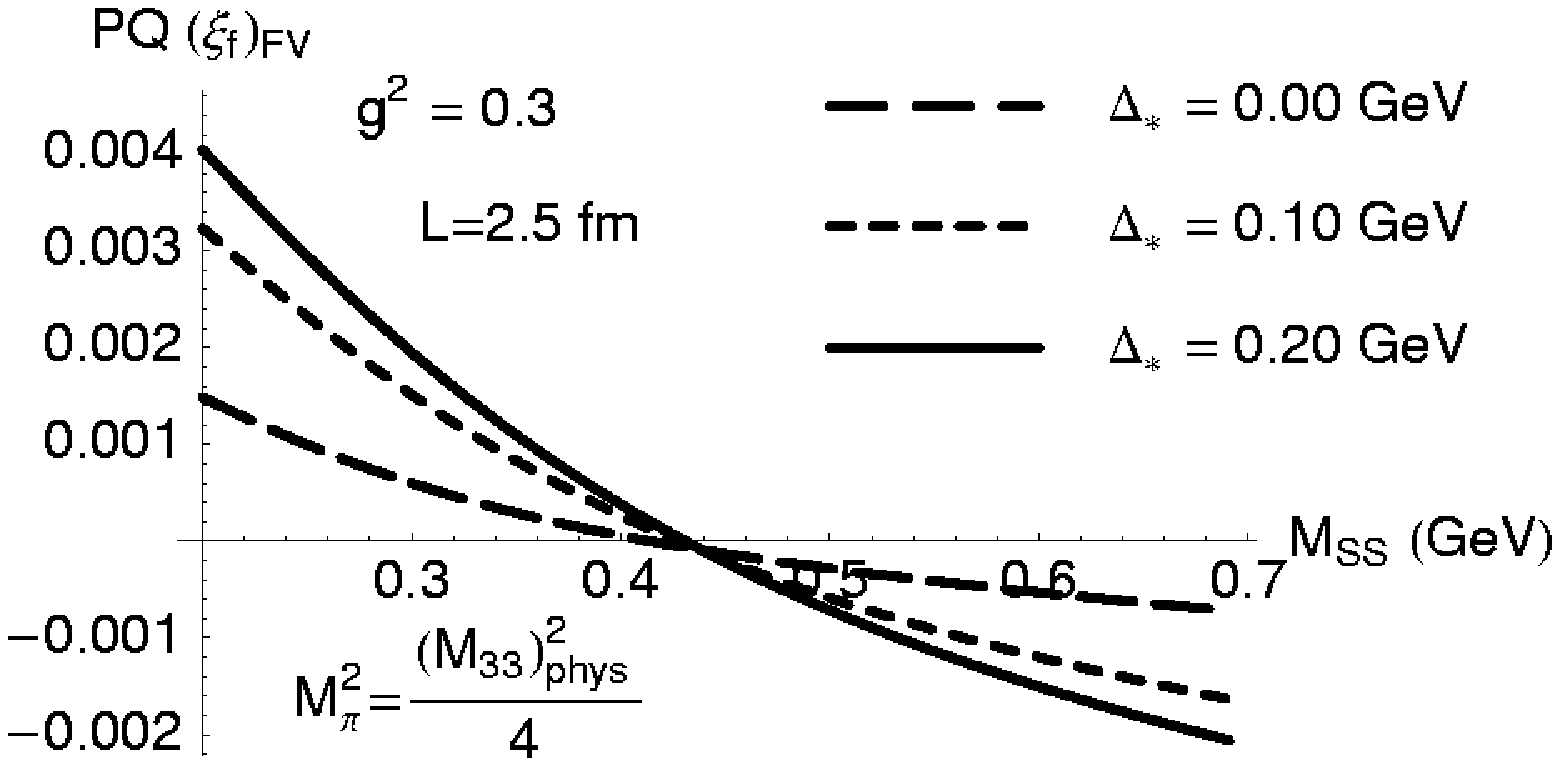}
\includegraphics[width=8.5cm]{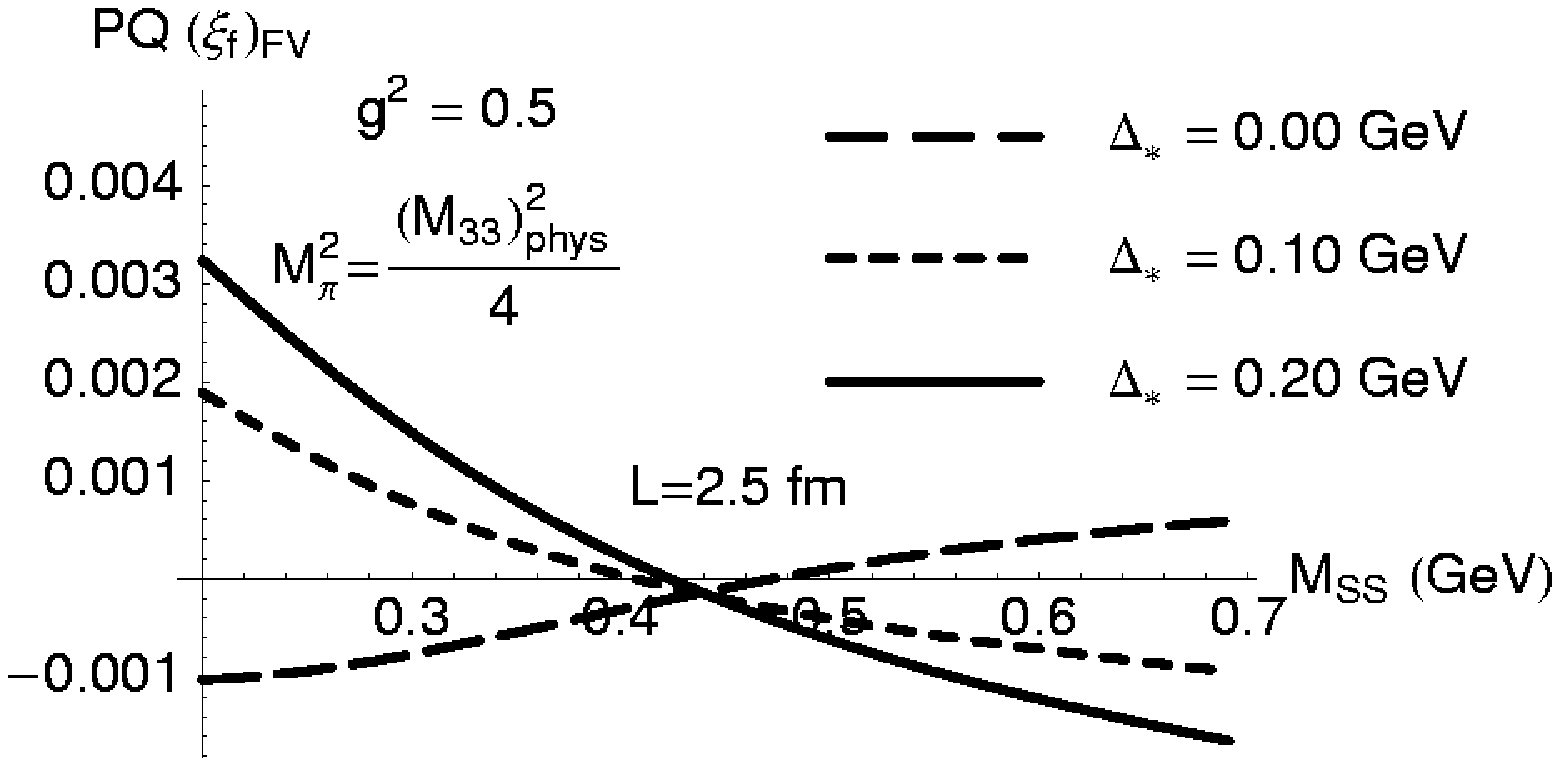}
\includegraphics[width=8.5cm]{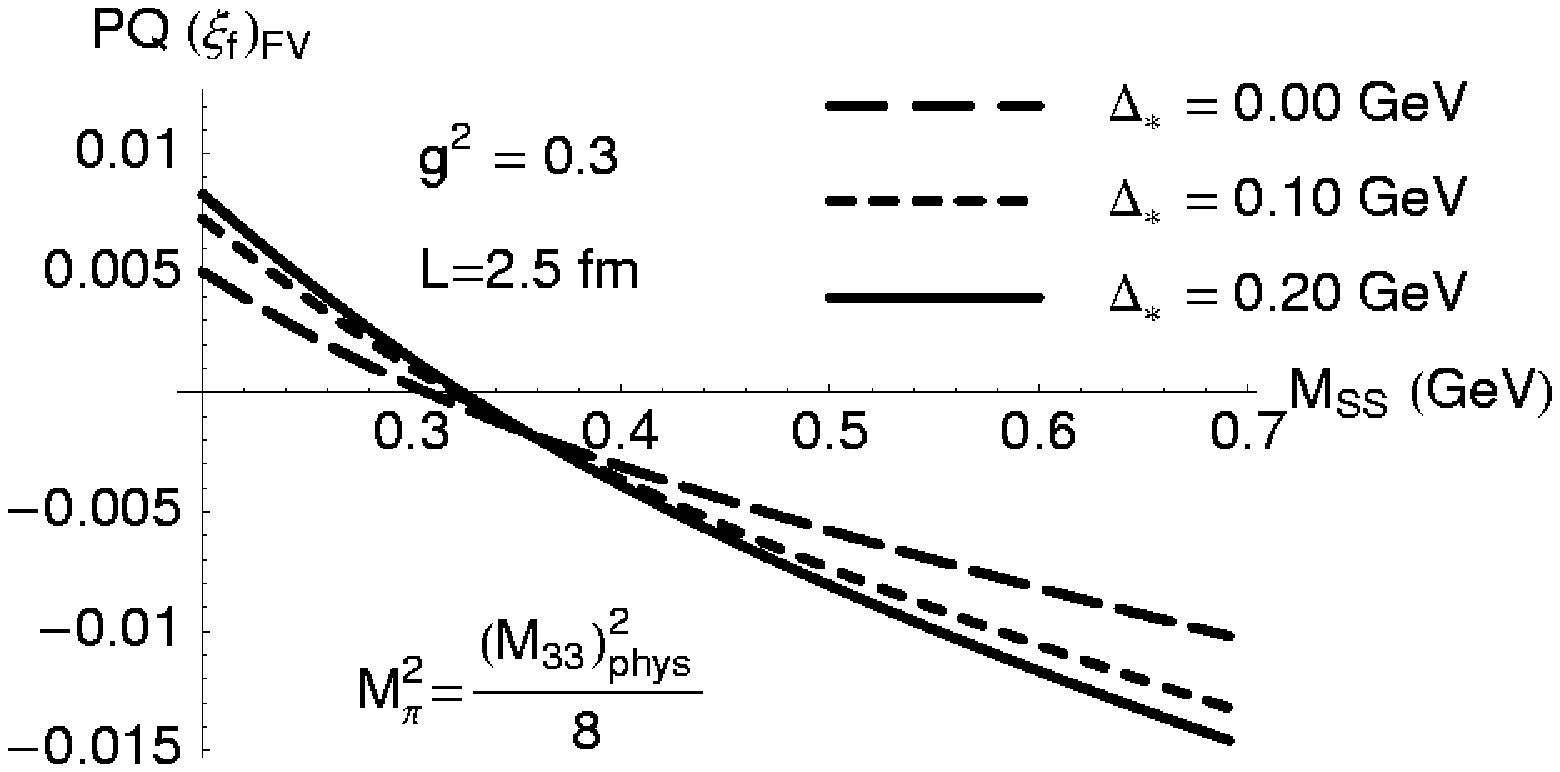}
\includegraphics[width=8.5cm]{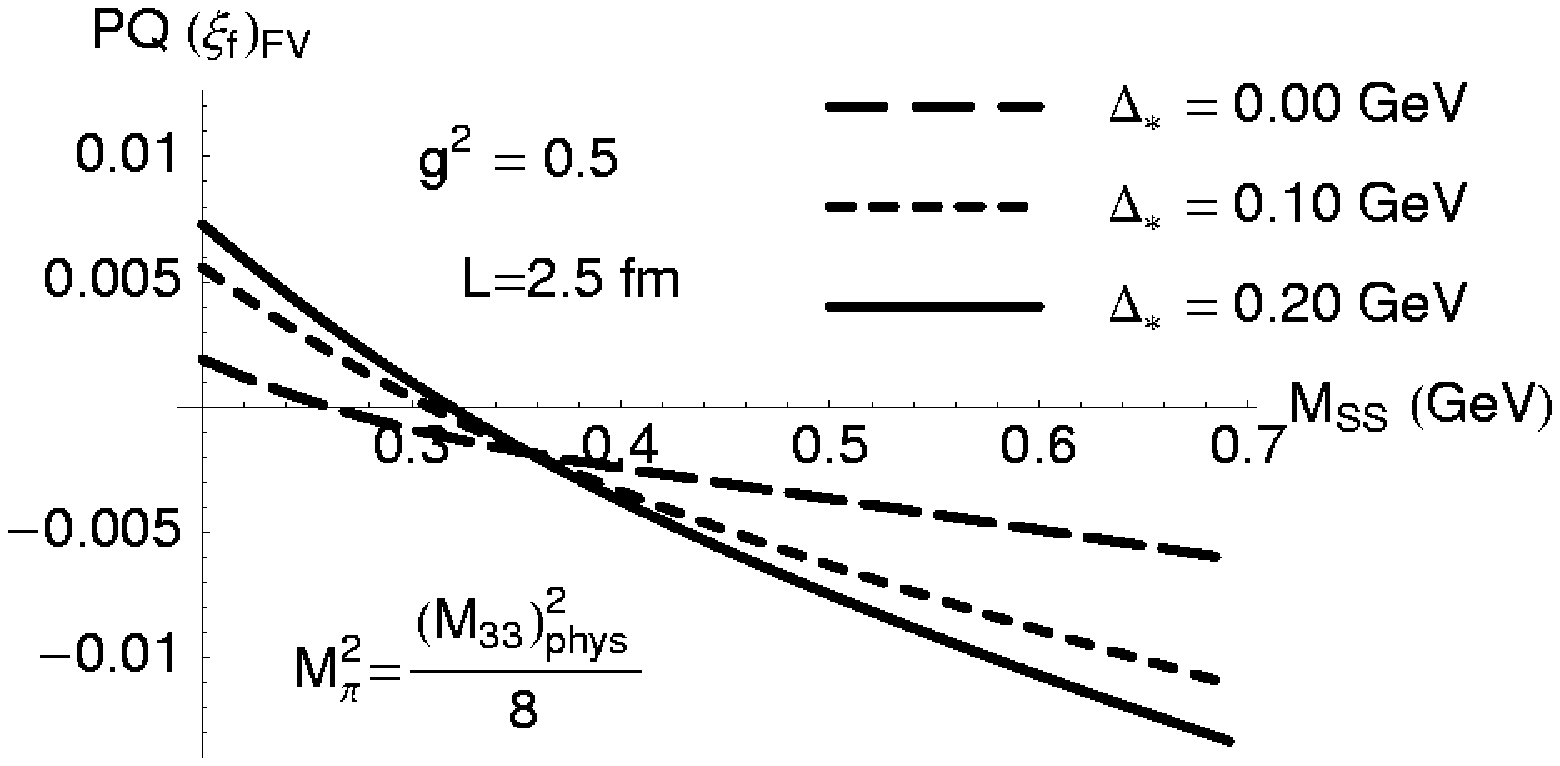}
\caption{\label{fig:PQ_fBs_over_fB}$(\xi_{f})_{\mathrm{FV}}$ 
in PQQCD plotted against
$M_{SS}$ (see text for the definition of $M_{SS}$), with $L=2.5$ fm
and two different values for $M_{\pi}$. 
The pion mass $M^{2}_{\pi}=M^{2}_{33}/4$ corresponds to 
$M_{\pi} L=4.4$ and $M^{2}_{\pi}=M^{2}_{33}/8$ corresponds to 
$M_{\pi} L=3.1$.
The mass $M_{SS} = 0.197$ GeV corresponds
to $M_{SS} L = 2.5$, and $M_{SS} = 0.32$ GeV corresponds
to $M_{SS} L = 4$ in this plot.}
\end{figure}
\begin{figure}
\includegraphics[width=8.5cm]{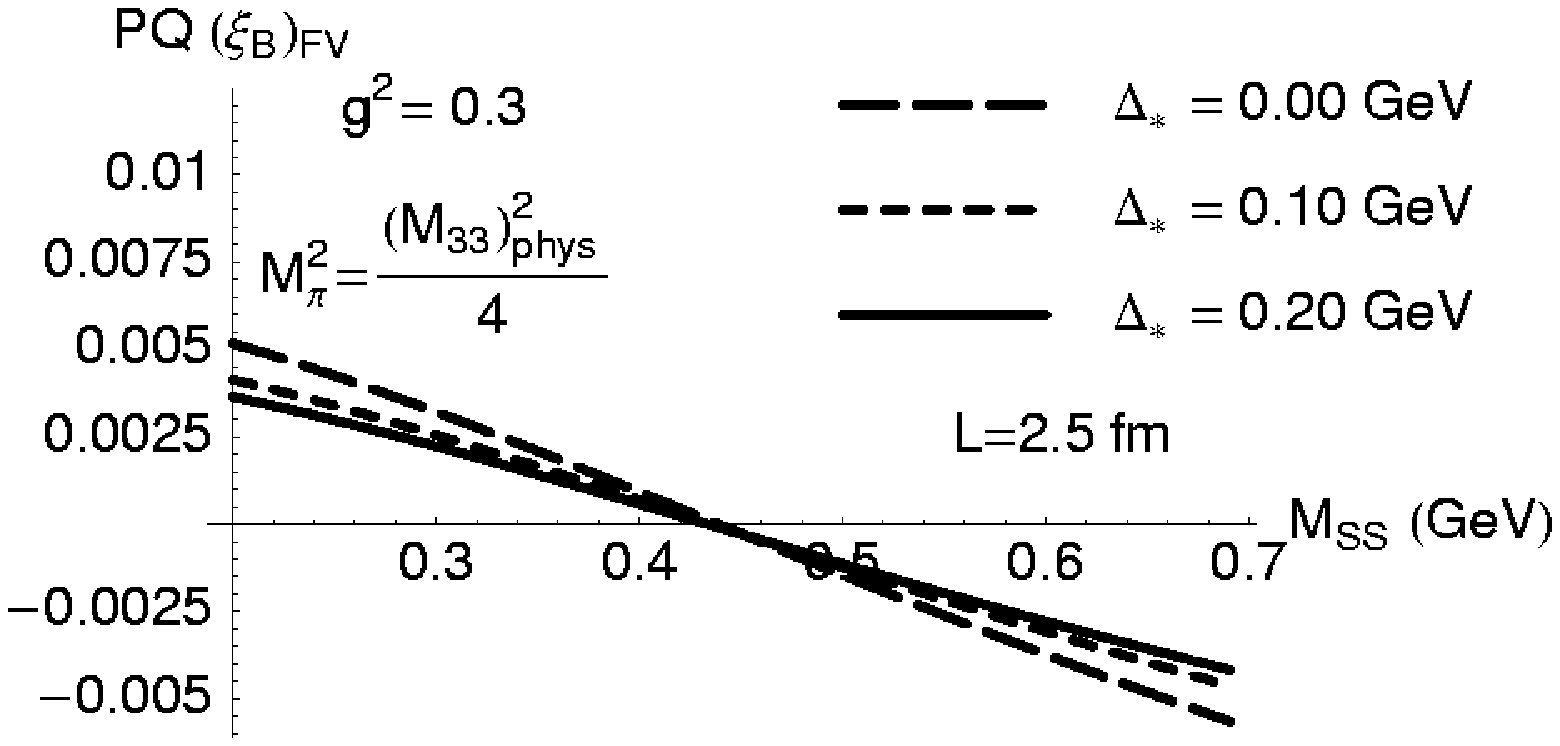}
\includegraphics[width=8.5cm]{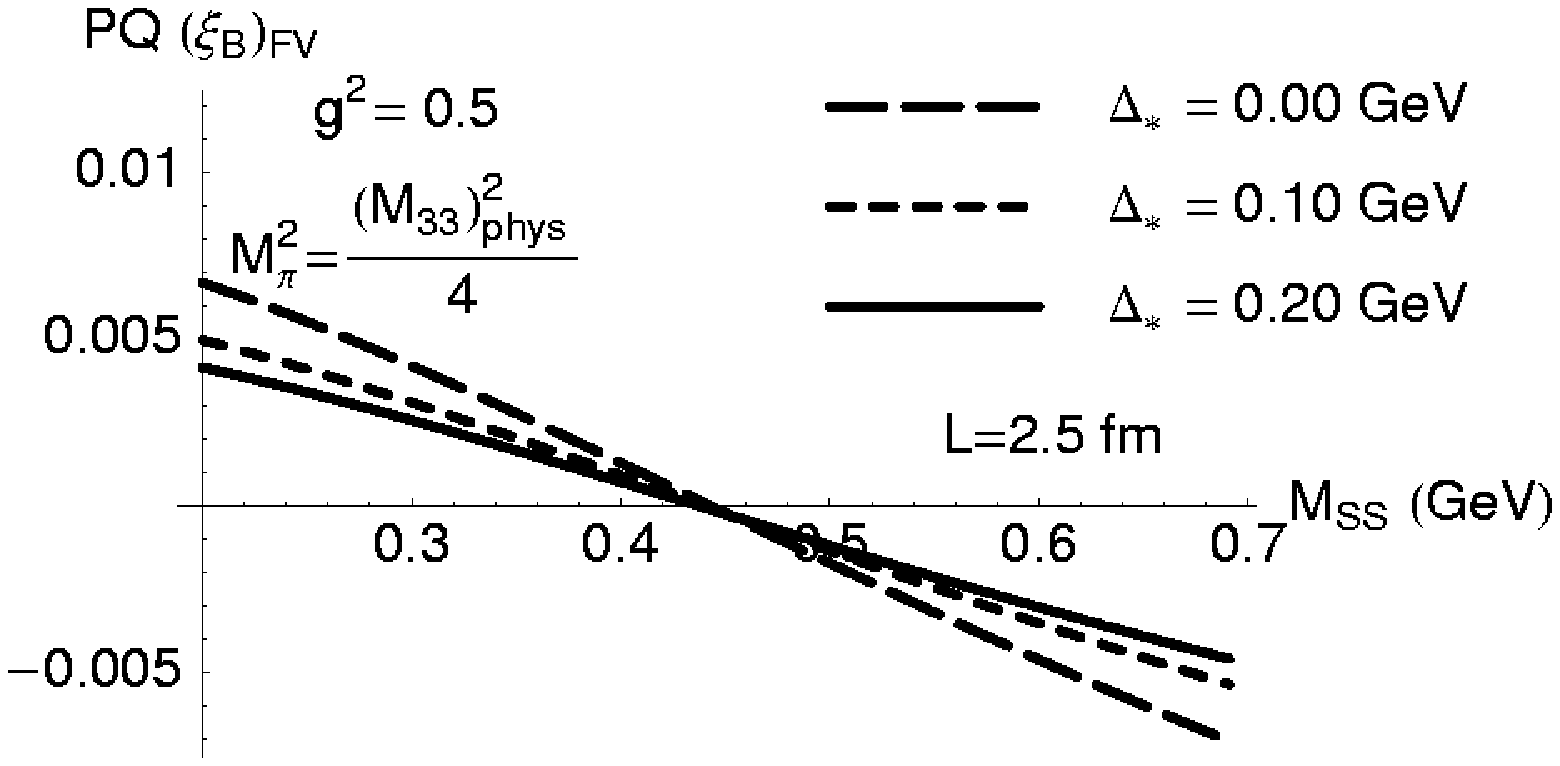}
\includegraphics[width=8.5cm]{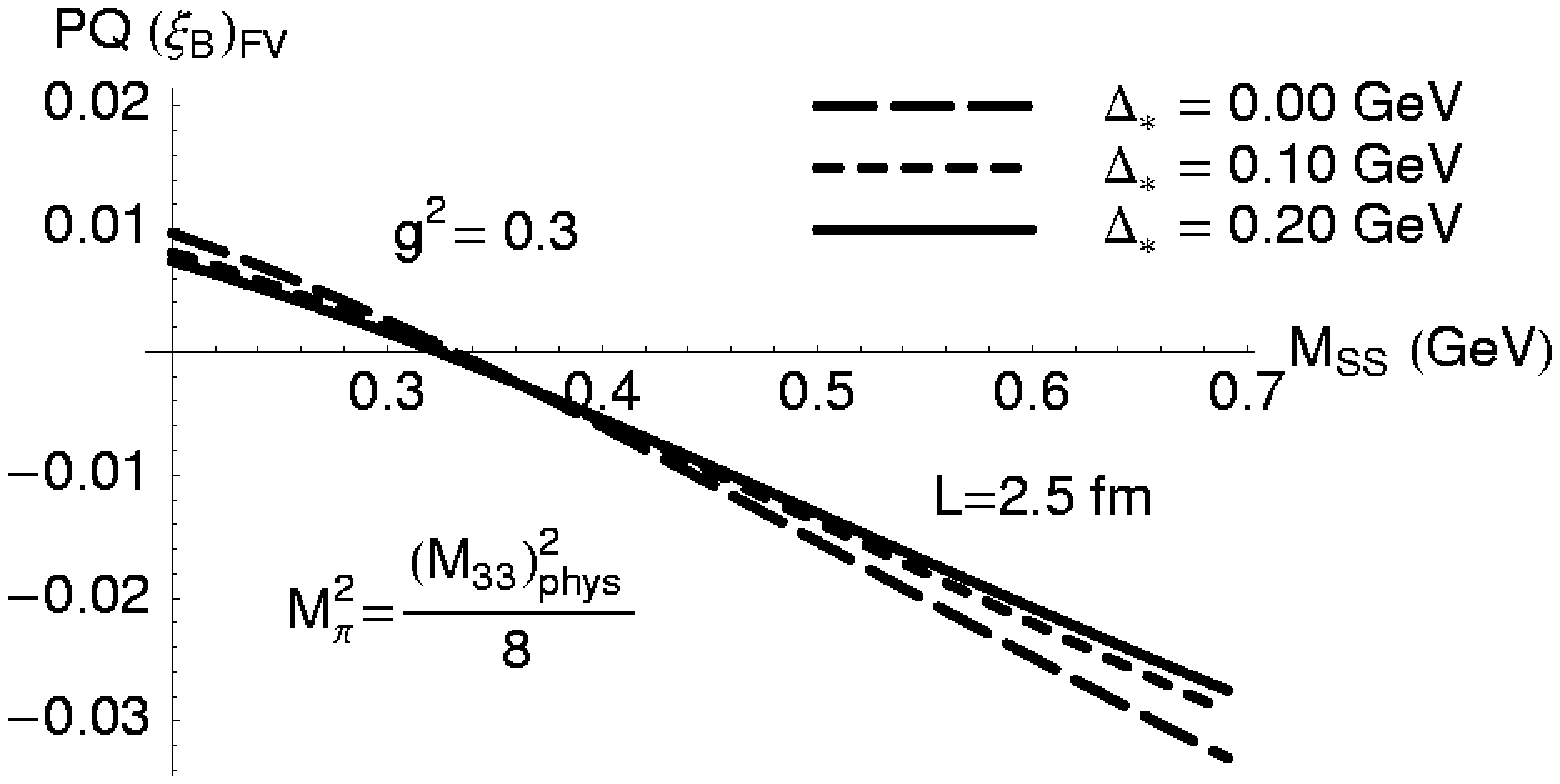}
\includegraphics[width=8.5cm]{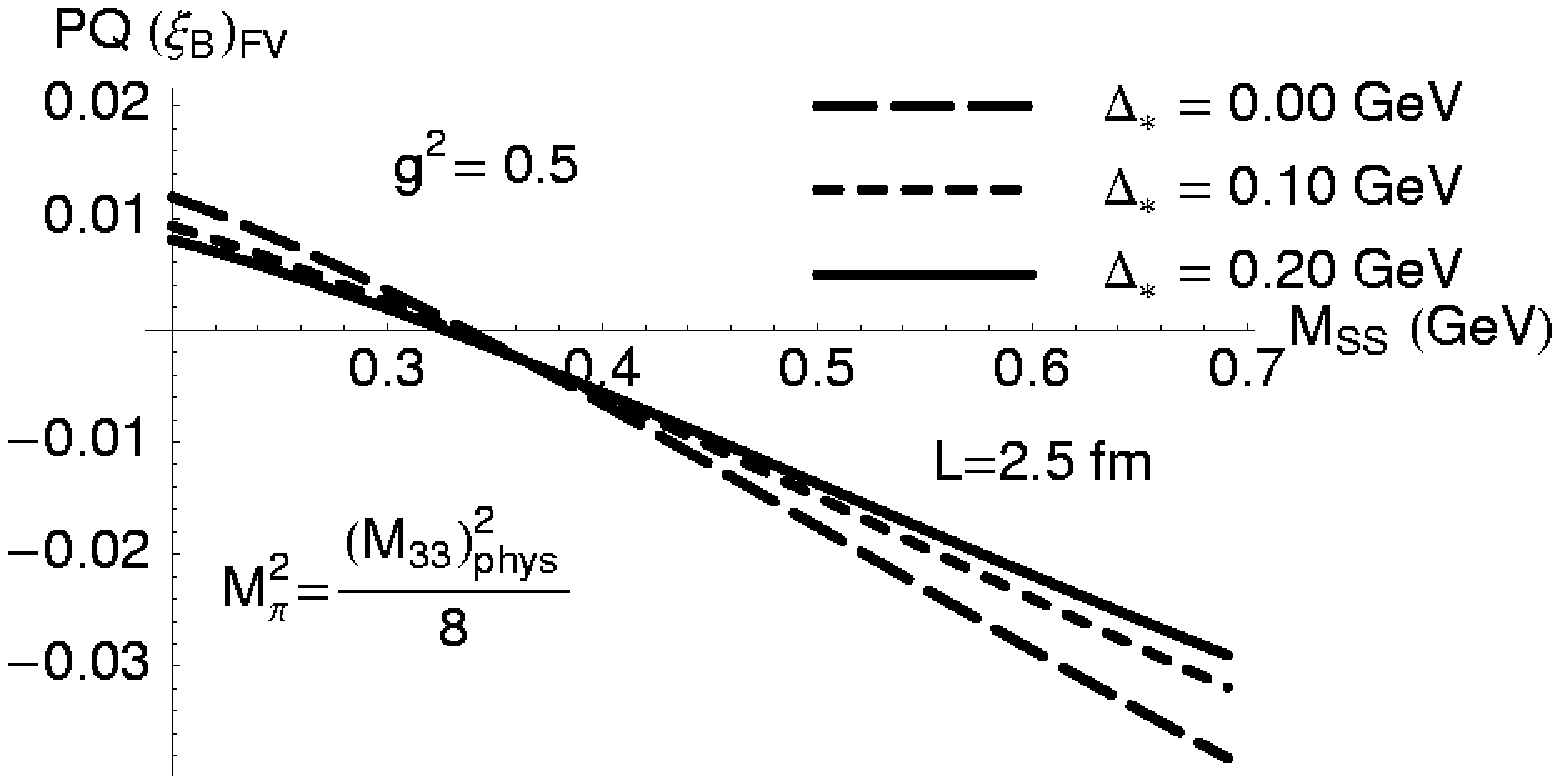}
\caption{\label{fig:PQ_BBs_over_BB}$(\xi_{B})_{\mathrm{FV}}$ 
in PQQCD plotted against
$M_{SS}$ (see text for the definition of $M_{SS}$), with $L=2.5$ fm
and two different values for $M_{\pi}$.
The pion mass $M^{2}_{\pi}=M^{2}_{33}/4$ corresponds to 
$M_{\pi} L=4.4$ and $M^{2}_{\pi}=M^{2}_{33}/8$ corresponds to 
$M_{\pi} L=3.1$. 
The mass $M_{SS} = 0.197$ GeV corresponds
to $M_{SS} L = 2.5$, and $M_{SS} = 0.32$ GeV corresponds
to $M_{SS} L = 4$ in this plot.}
\end{figure}
%
%
The results for the PQQCD analysis are presented in Figs.
\ref{fig:PQ_fBs_over_fB} and \ref{fig:PQ_BBs_over_BB}.  The double pole
insertions also appear in PQQCD and it is clear
from these plots that finite volume effects cannot be neglected 
if one hopes to determine $\xi$ to the level of a few percent.  Especially,
in the range of $M_{\pi}$ and $L$ where current and future 
lattice simulations are performed \cite{Davies:2003ik},  
they can already be at about 
$4\%$, and the dependence on the heavy meson mass is quite strong.
Therefore they can become comparable
to the error 
presented in the latest review~\cite{Kronfeld:2003sd},
\beq
 \xi = 1.23 \pm 0.10 
\eeq
after quark mass extrapolations.\footnote{Finite volume effects
presented in this work are, however, correlated with the errors arising
from chiral extrapolations.}

\section{Conclusion}
\label{sec:conclusion}
We have investigated finite volume effects in heavy quark systems in the
framework of heavy meson chiral perturbation theory.  The primary
conclusion
is that the scales $\Delta_{\ast}$ and $\delta_{s}$, which are 
heavy-light meson mass
splittings arising from the breaking of heavy quark spin and light flavour
$SU(3)$ symmetries, 
can significantly reduce the
volume effects in diagrams involving heavy meson propagators in the loop.  
The physical picture of this phenomenon is that some heavy-light mesons
are off-shell in the effective theory, as a consequence of the velocity
superselection rule, with the virtuality characterised by the mass splittings.
The time uncertainty conjugate to this virtuality limits the 
period during which the Goldstone particles can propagate to the
boundary.  Finite volume effects caused by the propagation of the
Goldstone particles naturally affect the light quark mass 
extrapolation/interpolation in a lattice calculation.  On top of this,
our work implies that they also influence the heavy quark
mass extrapolation/interpolation, since the scale $\Delta_{\ast}$ varies
significantly with the heavy meson mass.  The strength of this influence is
process-dependent, determined also by the relative weight between 
diagrams with and without heavy meson propagators in the loop.
The volume effects can
be amplified by both heavy and light quark mass extrapolations.
Therefore it is important
to perform calculations to identify these effects in phenomenologically 
interesting quantities.

%
%

We have presented an 
explicit calculation in finite volume HM$\chi$PT for the $B$ parameters in
neutral $B$ meson mixing and heavy-light decay constants, in full, quenched,
and partially quenched QCD. We have used these results to estimate the impact
of finite volume effects in the $SU(3)$ ratio $\xi$, which is an important
input in determining the magnitude of the CKM matrix element $V_{td}$.
Within the parameter space where most quenched lattice calculations have been
performed, we find that, although this impact is quite small
($\le \sim 2\%$) in full QCD, it can be significant in QQCD.
This is due to the enhanced long-distance effects arising
from the double pole structure.  This error will be amplified by
the quark mass extrapolations and hence can exceed the currently quoted
systematic effects.  Furthermore, finite volume effects tend towards
different directions 
in full and quenched QCD for decreasing $M_{\pi}$.  
This means that quenching errors in $\xi$
may be significantly larger than what was estimated before.
Therefore one has to be cautious when
using the existing quenched lattice QCD results for $\xi$ in phenomenological
work.  In
PQQCD, our results indicate that finite volume effects are typically
between $3\%$ and $5\%$ in the data range 
of future high-precision simulations, and they can be significantly
amplified in the procedure of quark mass extrapolations.
This means that they are not negligible in future lattice calculations of $\xi$.

\section*{Acknowledgments}
We warmly thank Silas Beane, Will Detmold, Laurent Lellouch, Martin Savage, 
Steve Sharpe and Ruth Van de Water for many helpful discussions.
The authors acknowledge the US Department of Energy grant
DE-FG03-97ER41014.
CJDL is also supported by grants DE-FG03-00ER41132 and DE-FG03-96ER40956.

\appendix
\section{Integrals and sums}
\label{app:integral}
We have regularised ultra-violet divergences that appear in loop integrals 
using dimensional
regularisation, and subtracted the term
\beq
\label{eq:lambdabar}
 \bar{\lambda} = \frac{2}{4 - d} - \gamma_{E} + {\mathrm{log}}(4\pi) + 1.
\eeq
The integrals appearing in the full QCD calculation are defined by 
\bea
 I_{\bar{\lambda}}(m) &\equiv&
 \mu^{4 - d}\int \frac{d^{d}k}{(2\pi)^{d}} \frac{1}{k^{2}-m^{2}+ i\epsilon}
\nonumber\\
   &=& \frac{i m^{2}}{16 \pi^{2}}
    \left [ \bar{\lambda} - {\log\left ( \frac{m^{2}}{\mu^{2}}\right )}
     \right ] , 
\eea
\bea
 H_{\bar{\lambda}}(m,\Delta) &\equiv&
 \left ( g^{\rho\nu} - v^{\rho} v^{\nu}\right ) \mu^{4 - d}  
\nonumber\\&&\times
\frac{\partial}{\partial\Delta}\int
 \frac{d^{d}k}{(2\pi)^{d}} 
  \frac{k_{\rho} k_{\nu}}{(k^{2}-m^{2}
+ i\epsilon)(v\cdot k - \Delta + i\epsilon)}
\nonumber\\
 &=& 3 \frac{\partial}{\partial\Delta} F_{\bar{\lambda}}(m,\Delta),
\eea
where
%
\beq
\label{eq:FandR}
 F_{\bar{\lambda}}(m,\Delta) = \frac{i}{16\pi^{2}} \bigg \{ 
 \left [ \bar{\lambda} - {\log\left ( \frac{m^{2}}{\mu^{2}}\right )}
     \right ] \left ( \frac{2\Delta^{2}}{3} - m^{2}\right )\Delta
   + \left ( \frac{10\Delta^{2}}{9} - \frac{4 m^{2}}{3}\right ) \Delta
   + \frac{2 (\Delta^{2}-m^{2})}{3} m R\left ( \frac{\Delta}{m}\right )
\bigg \} ,
\eeq
%
with
\beq
\label{eq:FandR_again}
R(x) \equiv \sqrt{x^{2}-1}\mbox{ }{\mathrm{log}}\left ( 
 \frac{x - \sqrt{x^{2}-1+i\epsilon}}{x + \sqrt{x^{2}-1+i\epsilon}} 
\right ) ,
\eeq
and $\mu$ is the renormalisation scale.
For the quenched and partially quenched calculations, we also need the 
integrals
\beq
 I^{(\eta^{\prime})}_{\bar{\lambda}} \equiv
 \mu^{4 - d}\int \frac{d^{d}k}{(2\pi)^{d}} 
\frac{1}{(k^{2}-m^{2}+ i\epsilon)^{2}}
   = \frac{\partial I_{\bar{\lambda}}(m)}{\partial m^{2}} , 
\eeq
and
\bea
 H^{\eta^{\prime}}_{\bar{\lambda}}(m,\Delta) 
&\equiv& 
 \left ( g^{\rho\nu} - v^{\rho} v^{\nu}\right ) \mu^{4 - d} 
\nonumber\\
&&\times
 \frac{\partial}{\partial\Delta}\int
 \frac{d^{d}k}{(2\pi)^{d}} 
  \frac{k_{\rho} k_{\nu}}{(k^{2}-m^{2}+i\epsilon)^{2}
  (v\cdot k - \Delta + i\epsilon)}
\nonumber\\
 &=& \frac{\partial}{\partial m^{2}} H_{\bar{\lambda}}(m,\Delta).
\eea
In a cubic spatial box of size $L$ in four dimensions with 
periodic boundary condition, so that the three-momentum is quantised
as in Eq.~(\ref{eq:k_quantum}),
one obtains the sums (after subtracting
the ultra-violet divergences)
\beq
\label{eq:FV_I}
 {\mathcal{I}}(m) 
\equiv \frac{1}{L^{3}}\sum_{\vec{k}}\int \frac{d k_{0}}{2\pi}
  \frac{1}{k^{2}-m^{2}+i\epsilon} = I(m) + I_{{\mathrm{FV}}}(m),
\eeq
and
%
\beq
 {\mathcal{H}}(m,\Delta) \equiv
 \left ( g^{\rho\nu} - v^{\rho} v^{\nu}\right )\left ( \frac{1}{L^{3}}\right ) 
 \sum_{\vec{k}} 
\frac{\partial}{\partial\Delta}\int
 \frac{d k_{0}}{2\pi} 
  \frac{k_{\rho} k_{\nu}}{(k^{2}-m^{2}
+i\epsilon)(v\cdot k - \Delta + i\epsilon)}
\label{eq:FV_H}
 = H(m,\Delta) + H_{{\mathrm{FV}}}(m,\Delta)
\eeq
for the full QCD calculation, where
\beq
 I(m) = I_{\bar{\lambda}}(m)|_{\bar{\lambda}=0} ,
\eeq
and
\beq
 H(m) = H_{\bar{\lambda}}(m,\Delta)|_{\bar{\lambda}=0} ,
\eeq
are the infinite volume limits of ${\mathcal{I}}$ and ${\mathcal{H}}$,
and ($n=|\vec{n}|$)
\bea
 I_{\mathrm{FV}}(m) 
  &=& \frac{-i}{4 \pi^{2}} m \sum_{\vec{n}\not=\vec{0}}
 \frac{1}{n L} K_{1}\left (n m L\right )\nonumber\\
& &\nonumber\\
 &\stackrel{m L \gg 1}{\longrightarrow}&
 \frac{-i}{4\pi^{2}} \sum_{\vec{n}\not=\vec{0}}
 \sqrt{\frac{m\pi}{2 n L}}
  \left (\frac{1}{n L}\right ) \exponential^{-n m L}
  \times \left \{
  1 + \frac{3}{8 n m L} 
 - \frac{15}{128 (n m L)^{2}}
 + \op\left ( \left [\frac{1}{n m L}\right ]^{3}\right )
 \right \}
\eea
is the finite volume correction to $I(m)$.  The function $H_{\mathrm{FV}}$
is the finite volume correction to $H(m,\Delta)$ and can be obtained via
\beq
 H_{\mathrm{FV}}(m,\Delta) = i \bigg[ 
 (m^{2} - \Delta^{2}) K_{\mathrm{FV}}(m,\Delta) 
  - 2 \Delta J_{\mathrm{FV}}(m,\Delta)
  + i I_{\mathrm{FV}}(m)
 \bigg ] ,
\eeq
where $J_{\mathrm{FV}}(m,\Delta)$ and $K_{\mathrm{FV}}(m,\Delta)$ 
are defined in Eqs.~(\ref{eq:JFV}) and (\ref{eq:KFV_asymptotic}).

%
%

For QQCD and PQQCD calculations, one also needs
\beq
\label{eq:FV_I_DP}
 {\mathcal{I}}^{\eta^{\prime}}(m) \equiv \frac{1}{L^{3}}
   \sum_{\vec{k}}\int \frac{d k_{0}}{2\pi}
  \frac{1}{(k^{2}-m^{2}+i\epsilon)^{2}} = \frac{\partial I(m)}{\partial m^{2}}
 + \frac{\partial I_{\mathrm{FV}}(m)}{\partial m^{2}} ,
\eeq
and
\bea
 {\mathcal{H}}^{\eta^{\prime}}(m,\Delta) 
&\equiv& \frac{\partial}{\partial\Delta}
 \left [
 \left ( g^{\rho\nu} - v^{\rho} v^{\nu}\right )\left ( \frac{1}{L^{3}}\right ) 
 \sum_{\vec{k}} \int
 \frac{d k_{0}}{2\pi} 
  \frac{k_{\rho} k_{\nu}}{(k^{2}-m^{2}
+i\epsilon)^{2}(v\cdot k - \Delta + i\epsilon)}
 \right ] \nonumber\\
& &\nonumber\\
\label{eq:FV_H_DP}
 &=& \frac{\partial H(m,\Delta)}{\partial m^{2}} + 
  \frac{\partial H_{\mathrm{FV}}(m,\Delta)}{\partial m^{2}}.
\eea

\section{One-loop results for decay constants and $B$ parameters}
\label{sec:results}
In this appendix, we collect the results for one-loop corrections to 
$f_{P_{(s)}}\sqrt{M_{P_{(s)}}}$ and
$B_{P_{(s)}}$.    For convenience, we introduce
\beq
 \calC_{\pm}(M_{\mathrm{GP}},x) = {\mathcal{I}}(M_{\mathrm{GP}})
      \pm g^{2} {\mathcal{H}}(M_{\mathrm{GP}},x) ,
\eeq
and
\beq
 \calC^{\etaprime}_{\pm}(M_{\mathrm{GP}},x) =
  {\mathcal{I}}^{\etaprime}(M_{\mathrm{GP}})
      \pm g^{2} {\mathcal{H}}^{\etaprime}(M_{\mathrm{GP}},x) ,
\eeq
where the functions ${\mathcal{I}}(m)$, ${\mathcal{H}}(m,x)$, 
${\mathcal{I}}^{\eta^{\prime}}(m)$ and 
${\mathcal{H}}^{\eta^{\prime}}(m,x)$ 
are defined in Eqs.
(\ref{eq:FV_I}), (\ref{eq:FV_H}), (\ref{eq:FV_I_DP}) and (\ref{eq:FV_H_DP}), 
respectively.

%
%

In full QCD, we find
\bea
 f_{P}\sqrt{M_{P}} &=& \kappa
 \bigg \{
  1 - \frac{i}{12 f^{2}} \bigg [ 
   9\mbox{ }\calC_{-}(M_{\pi},\Delta_{\ast}) 
   + 6\mbox{ }\calC_{-}(M_{K},\Delta_{\ast}+\delta_{s})
   + \calC_{-}(M_{\eta},\Delta_{\ast})
 \bigg ]
 \bigg \} ,\\
 & &\nonumber\\
f_{P_{s}}\sqrt{M_{P_{s}}} &=& \kappa
 \left \{
  1 - \frac{i}{3 f^{2}} \bigg [
    3\mbox{ }\calC_{-}(M_{K},\Delta_{\ast}-\delta_{s}) 
   + \calC_{-}(M_{\eta},\Delta_{\ast}) 
 \bigg ]
 \right \} ,\\
& &\nonumber\\
 B_{P} &=& \frac{3\beta}{2\kappa^{2}}
 \left \{
  1 - \frac{i}{6 f^{2}} \bigg [ 
   3\mbox{ }\calC_{+}(M_{\pi},\Delta_{\ast}) 
   + \calC_{+}(M_{\eta},\Delta_{\ast})
 \bigg ]
 \right \} ,\nonumber\\
 & &\\
 B_{P_{s}} &=& \frac{3\beta}{2\kappa^{2}}
 \left \{
  1 - \frac{2 i}{3 f^{2}} \bigg [
   \calC_{+}(M_{\eta},\Delta_{\ast})
 \bigg ]
 \right \} .
\eea
In QQCD, we find
\bea
f_{P}\sqrt{M_{P}} &=& \kappa\bigg \{
 1 + \frac{i}{2 f^{2}} \bigg [
 \frac{\alpha}3\calC_{-}(M_{\pi},\Delta_{\ast})
 + \left (\frac{\alpha M^{2}_{\pi} - M^{2}_{0}}{3}\right )
 \calC^{\etaprime}_{-}(M_{\pi},\Delta_{\ast})
 + 2g\gamma\zsum(M_{\pi},\Delta_{\ast})
\bigg ]
\bigg \} ,\\
& &\nonumber\\
f_{P_{s}}\sqrt{M_{P_{s}}} &=& \kappa\bigg \{
 1 + \frac{i}{2 f^{2}} \bigg [
 \frac{\alpha}{3}\calC_{-}(M_{33},\Delta_{\ast})
 + \left (\frac{\alpha M^{2}_{33} - M^{2}_{0}}{3}\right )
 \calC^{\etaprime}_{-}(M_{33},\Delta_{\ast})
 + 2g\gamma\zsum(M_{33},\Delta_{\ast})
\bigg ]
\bigg \} ,\\
& &\nonumber\\
B_{P} &=& \frac{3\beta}{2\kappa^{2}} \bigg \{
 1 - \frac{i}{f^{2}} \bigg [
 \left (1 - \frac{\alpha}{3}\right )\calC_{+}(M_{\pi},\Delta_{\ast})
 - \left (\frac{\alpha M^{2}_{\pi} - M^{2}_{0}}{3}\right )
\calC^{\etaprime}_{+}(M_{\pi},\Delta_{\ast})
 + 2g\gamma\zsum(M_{\pi},\Delta_{\ast})
\bigg ]
\bigg \} ,\\
& &\nonumber\\
B_{P_{s}} &=& \frac{3\beta}{2\kappa^{2}}\bigg \{
 1 - \frac{i}{f^{2}} \bigg [
 \left (1 - \frac{\alpha}{3}\right )\calC_{+}(M_{33},\Delta_{\ast})
 - \left (\frac{\alpha M^{2}_{33} - M^{2}_{0}}{3}\right )
\calC^{\etaprime}_{+}(M_{33},\Delta_{\ast})
 + 2g\gamma\zsum(M_{33},\Delta_{\ast})
\bigg ]
\bigg \} ,
\eea
where
\beq
\label{eq:m33}
 M_{33} = \sqrt{ 2 M^{2}_{K} - M_{\pi}^{2} }.
\eeq
In PQQCD, we find
\bea
 f_{P}\sqrt{M_{P}} &=& \kappa\bigg \{ 1 - \frac{i}{12 f^{2}}
 \bigg [
  12\mbox{ }\calC_{-}(M_{VS},\Delta_{\ast}+\delta_{\mathrm{sea}})
  + 6\mbox{ }\calC_{-}(M_{\tau}, 
    \Delta_{\ast}+\delta_{\mathrm{sea}}+\tilde{\delta}_{s})
\nonumber\\
 & &\mbox{ }\mbox{ }\mbox{ }\mbox{ }\mbox{ }\mbox{ }\mbox{ }\mbox{ }
  + 3
     \frac{M^{2}_{33} - M^{2}_{\pi} - 2\delta^{2}_{VSs}}
     {M^{2}_{\pi}-M^{2}_{33}+\delta^{2}_{VS}+2\delta^{2}_{VSs}}
    \calC_{-}(M_{\pi},\Delta_{\ast})
  + \left (\frac{M_{\pi}^{2}-M_{33}^{2}-2\delta^{2}_{VS}+2\delta^{2}_{VSs}}
  {M_{\pi}^{2}-M_{33}^{2}
      +\delta^{2}_{VS}+2\delta^{2}_{VSs}}\right )^{2}
   \calC_{-}(M_{X},\Delta_{\ast})\nonumber\\
 & &\mbox{ }\mbox{ }\mbox{ }\mbox{ }\mbox{ }\mbox{ }\mbox{ }\mbox{ }
    \mbox{ }\mbox{ }\mbox{ }\mbox{ }\mbox{ }\mbox{ }\mbox{ }\mbox{ }
    \mbox{ }\mbox{ }
  + 6\frac{\delta^{2}_{VS}(M_{33}^{2}-M_{\pi}^{2}-2\delta^{2}_{VSs})}{M_{\pi}^{2}
  -M_{33}^{2}+\delta^{2}_{VS}+2\delta^{2}_{VSs}}
\calC^{\etaprime}_{-}(M_{\pi},\Delta_{\ast})
 \bigg ]
 \bigg \} , \\
& &\nonumber\\
& &\nonumber\\
 f_{P_{s}}\sqrt{M_{P_{s}}} &=& \kappa
\bigg \{ 1 - \frac{i}{6 f^{2}}\bigg [
  3\mbox{ }\calC_{-}(M_{VSs},
  \Delta_{\ast}+\delta_{\mathrm{sea}}+\tilde{\delta}_{s}-\delta_{s})
  + 6\mbox{ }\calC_{-}(M_{\zeta s}, 
  \Delta_{\ast}+\delta_{\mathrm{sea}}-\delta_{s})\nonumber\\
 & &\mbox{ }\mbox{ }\mbox{ }\mbox{ }
  -3
  \frac{M^{2}_{33}-M^{2}_{\pi}+2\delta^{2}_{VS}}
       {M^{2}_{33}-M^{2}_{\pi}+2\delta^{2}_{VS}+4\delta^{2}_{VSs}}
   \calC_{-}(M_{33},\Delta_{\ast})
  + 2\left (\frac{M_{\pi}^{2}-M_{33}^{2}
       -2\delta^{2}_{VS}+2\delta^{2}_{VSs}}
  {M_{\pi}^{2}-M_{33}^{2}
       -2\delta^{2}_{VS}-4\delta^{2}_{VSs}}\right )^{2}
   \calC_{-}(M_{X},\Delta_{\ast})\nonumber\\
 & &\mbox{ }\mbox{ }\mbox{ }\mbox{ }\mbox{ }\mbox{ }\mbox{ }\mbox{ }
    \mbox{ }\mbox{ }\mbox{ }\mbox{ }\mbox{ }\mbox{ }\mbox{ }\mbox{ }
    \mbox{ }\mbox{ }\mbox{ }\mbox{ }
  + 6\frac{\delta^{2}_{VSs}(M_{33}^{2}-M_{\pi}^{2}+2\delta^{2}_{VS})}
{M_{\pi}^{2}-M_{33}^{2}
  -2\delta^{2}_{VS}-4\delta^{2}_{VSs}}
  \calC^{\etaprime}_{-}(M_{33},\Delta_{\ast})
 \bigg ]
 \bigg \} , \\
& &\nonumber\\
& &\nonumber\\
 B_{P} &=& \frac{3\beta}{2\kappa^{2}}
\bigg \{ 1 - \frac{i}{6 f^{2}}\bigg [
   6\mbox{ }\calC_{+}(M_{\pi},\Delta_{\ast})
  - 3\frac{M^{2}_{\pi}-M^{2}_{33}+2\delta^{2}_{VSs}}
     {M^{2}_{\pi}-M^{2}_{33}+\delta^{2}_{VS}+2\delta^{2}_{VSs}}
    \calC_{+}(M_{\pi},\Delta_{\ast})
\nonumber\\
 & &\mbox{ }\mbox{ }\mbox{ }\mbox{ }\mbox{ }\mbox{ }\mbox{ }\mbox{ }
    \mbox{ }\mbox{ }\mbox{ }\mbox{ }\mbox{ }\mbox{ }\mbox{ }\mbox{ }
    \mbox{ }\mbox{ }\mbox{ }\mbox{ }
  + \left (\frac{M_{\pi}^{2}-M_{33}^{2}-2\delta^{2}_{VS}+2\delta^{2}_{VSs}}
  {M_{\pi}^{2}-M_{33}^{2}+
   \delta^{2}_{VS}+2\delta^{2}_{VSs}}\right )^{2}
   \calC_{+}(M_{X},\Delta_{\ast})\nonumber\\
 & &\mbox{ }\mbox{ }\mbox{ }\mbox{ }\mbox{ }\mbox{ }\mbox{ }\mbox{ }
    \mbox{ }\mbox{ }\mbox{ }\mbox{ }\mbox{ }\mbox{ }\mbox{ }\mbox{ }
    \mbox{ }\mbox{ }\mbox{ }\mbox{ }
  +  
6\frac{\delta^{2}_{VS}(M_{33}^{2}-M_{\pi}^{2}-2\delta^{2}_{VSs})}
 {M_{\pi}^{2}-M_{33}^{2}+\delta^{2}_{VS}+2\delta^{2}_{VSs}}
 \calC^{\etaprime}_{+}(M_{\pi},\Delta_{\ast})
 \bigg ]
 \bigg \} , \\
& &\nonumber\\
& &\nonumber\\
 B_{P_{s}} &=& \frac{3\beta}{2\kappa^{2}}
  \bigg \{ 1 - \frac{i}{3 f^{2}}\bigg [
  3\calC_{+}(M_{33},\Delta_{\ast})  
-3\frac{M^{2}_{\pi}-M^{2}_{33}-2\delta^{2}_{VS}}
    {M^{2}_{\pi}-M^{2}_{33}-2\delta^{2}_{VS}-4\delta^{2}_{VSs}}
   \calC_{+}(M_{33},\Delta_{\ast})
\nonumber\\
 & &\mbox{ }\mbox{ }\mbox{ }\mbox{ }\mbox{ }\mbox{ }\mbox{ }\mbox{ }
    \mbox{ }\mbox{ }\mbox{ }\mbox{ }\mbox{ }\mbox{ }\mbox{ }\mbox{ }
    \mbox{ }\mbox{ }\mbox{ }\mbox{ }
  + 2\left (
  \frac{M_{\pi}^{2}-M_{33}^{2}-2\delta^{2}_{VS}+2\delta^{2}_{VSs}}
  {M_{\pi}^{2}-M_{33}^{2}-2\delta^{2}_{VS}-4\delta^{2}_{VSs}}
 \right )^{2}
   \calC_{+}(M_{X},\Delta_{\ast})\nonumber\\
 & &\mbox{ }\mbox{ }\mbox{ }\mbox{ }\mbox{ }\mbox{ }\mbox{ }\mbox{ }
    \mbox{ }\mbox{ }\mbox{ }\mbox{ }\mbox{ }\mbox{ }\mbox{ }\mbox{ }
    \mbox{ }\mbox{ }\mbox{ }\mbox{ }
  + 6\frac{\delta^{2}_{VSs}(M_{33}^{2}-M_{\pi}^{2}+2\delta^{2}_{VS})}
          {M_{\pi}^{2}-M_{33}^{2}-2
\delta^{2}_{VS}-4\delta^{2}_{VSs}}
 \calC^{\etaprime}_{+}(M_{33},\Delta_{\ast})
 \bigg ]
 \bigg \} ,
\eea
where
\bea
 M_{VS}^{2} = B_{0} (m + \tilde{m})&,&\mbox{ }\mbox{ }
 M_{VSs}^{2} = B_{0} (m_{s} + \tilde{m}_{s}) ,\nonumber\\
 M_{\tau}^{2}= B_{0} (m + \tilde{m}_{s}) = 
M^{2}_{VSs} - M^{2}_{K} + M^{2}_{\pi} &,&\mbox{ }\mbox{ }
 M_{\zeta s}^{2} = B_{0} (m_{s} + \tilde{m}) 
= M^{2}_{VS} + M^{2}_{K} - M^{2}_{\pi},\nonumber\\
 \delta^{2}_{VS} = M^{2}_{\pi} - M^{2}_{VS} &,&\mbox{ }\mbox{ }
 \delta^{2}_{VSs} = M^{2}_{33} - M^{2}_{VSs}, \nonumber\\
 {\mathrm{and}}\mbox{ }\mbox{ }
 M_{X}^{2} &=& \frac{1}{3}\left ( M^{2}_{\pi} + 2 M^{2}_{33}
     - 2 \delta^{2}_{VS} -4 \delta^{2}_{VSs}\right ).
\eea
It is straightforward to show that the PQQCD results 
reproduce those for full QCD in the limit 
$\tilde{m}=m$ and $\tilde{m}_{s}=m_{s}$.

\addcontentsline{toc}{chapter}{Bibliography} 
\bibliographystyle{prsty} 
\bibliography{refs} 
 
\end{document}